\documentclass[aps,reprint,twocolumn,superscriptaddress]{revtex4-2}
\usepackage{graphicx}
\usepackage{lipsum}
\usepackage{xr}
\usepackage{xcolor}
\usepackage{ulem}
\usepackage{enumitem}

\newcommand{\affekut}{\affiliation{%
Physikalisches Institut, Center for Quantum Science (CQ) and LISA$^+$,
Eberhard Karls Universit\"at T\"ubingen,
Auf der Morgenstelle 14,
72076 T\"ubingen, Germany}}

\newcommand{\affucsd}{\affiliation{%
Center for Advanced Nanoscience, Department of Physics, 
University of California –- San Diego, 
9500 Gilman Drive, La Jolla, CA92093-0319, USA}}

\newcommand{\afftech}{\affiliation{%
Department of Materials Science and Engineering, 
Technion -- Israel Institute of Technology, 
Technion City, 32000 Haifa, Israel}}

\newcommand{\affcnrs}{\affiliation{%
CNRS Laboratoire de Physique des Solides, 
Universit\'e Paris-Saclay, 
91405, Orsay, France}}

\newcommand{\afflorenzo}{\affiliation{%
CNRS Laboratoire de Physique Th\'eorique et Mod\'elisation,
CY Cergy Paris Universit\'e,
95302 Cergy-Pontoise Cedex, France}}

\begin{document}

\title{Laser-induced quenching of metastability at the Mott-insulator to metal transition}

\author{Theodor Luibrand}
%\email{theodor.luibrand@uni-tuebingen.de}
\affekut
\author{Lorenzo Fratino}
%\email{lorenzo.fratino@cyu.fr}
\afflorenzo
\affcnrs
\author{Farnaz Tahouni-Bonab}
%\email(farnaz.tahouni-bonab@uni-tuebingen.de)
\affekut
\author{Amihai Kronman}
%\email(amihaik@campus.technion.ac.il)
\afftech
\author{Yoav Kalcheim}
%\email{ykalcheim@technion.ac.il}
\afftech
\affucsd
\author{Ivan K.~Schuller}
%\email{ikssch@physics.ucsd.edu}
\affucsd
\author{Marcelo Rozenberg}
%\email{marcelo.rozenberg@universite-paris-saclay.fr}
\affcnrs
\author{Reinhold Kleiner}
%\email{kleiner@uni-tuebingen.de}
\affekut
\author{Dieter Koelle}
%\email{koelle@uni-tuebingen.de}
\affekut
\author{Stefan Gu\'enon}
\email{stefan.guenon@uni-tuebingen.de}
\affekut

\begin{abstract}
    There is growing interest in strongly correlated insulator thin films because the intricate interplay of their intrinsic and extrinsic state variables causes memristive behavior that might be used for bio-mimetic devices in the emerging field of neuromorphic computing.
    In this study we find that laser irradiation tends to drive V$_2$O$_3$ from supercooled/superheated metastable states towards thermodynamic equilibrium, most likely in a non-thermal way.
    We study thin films of the prototypical Mott-insulator V$_2$O$_3$, which show spontaneous phase separation into metal-insulator herringbone domains during the Mott transition.
    Here, we use low-temperature microscopy to investigate how these metal-insulator domains can be modified by scanning a focused laser beam across the thin film surface.
    We find that the response depends on the thermal history:   
    When the thin film is heated from below the Mott transition temperature, the laser beam predominantly induces metallic domains. 
    On the contrary, when the thin film is cooled from a temperature above the transition, the laser beam predominantly induces insulating domains. 
    Very likely, the V$_2$O$_3$ thin film is in a superheated or supercooled state, respectively, during the first-order phase transition, and the perturbation by a laser beam drives these metastable states into stable ones.
		This way, the thermal history is locally erased.
		Our findings are supported by a phenomenological model with a laser-induced lowering of the energy barrier between the metastable and equilibrium states.
\end{abstract}

\maketitle
\let\oldaddcontentsline\addcontentsline
\renewcommand{\addcontentsline}[3]{}

Strongly correlated insulator thin films are the subject of intensive research in the emerging field of neuromorphic computing \cite{schuller2015, valle2018, hoffmann2022, park2022}.
If these materials are cooled to a temperature below the transition temperature, the charge carrier mobility is considerably reduced due to electron-electron correlation effects like the Mott-Hubbard interaction (e.g., V$_2$O$_3$ \cite{morin1959, mcwhan1970, mcwhan1973, georges1996, limelette2003}), charge transfer (e.g., rare-earth nickelates \cite{torrance1992, catalano2018}) or dimerization (e.g., VO$_2$)\cite{biermann2005}.
This reduction in charge carrier mobility results in a metal-to-insulator transition (MIT) \cite{imada1998}. 
An equivalent insulator-to-metal transition (IMT) is observed during heating.
The electron-electron correlation effects cause an intricate interplay of intrinsic and extrinsic state variables that can be leveraged for engineering neuromorphic devices \cite{zhou2015}.
One way to influence strongly correlated insulator thin-film devices externally is through exposure to light.
For instance, light is used to trigger resistive switching \cite{kim2014, kim2018, seo2012, li2022a}, to tune the resistive switching voltage thresholds \cite{lee2012} or to tune the frequency of relaxation oscillators \cite{seo2012a} in VO$_2$ devices.
Furthermore, because strongly correlated insulator thin-film neuromorphic devices are usually operated in a critical state at the onset of the IMT/MIT, their functionality is affected by thermodynamics of phase transitions.
For instance, the hysteresis in resistance $R$ vs temperature $T$ curves due to the first order MIT/IMT in VO$_2$ is leveraged for multistate resistive switching \cite{driscoll2009, rana2020, gao2022}.
In V$_2$O$_3$, the MIT/IMT is accompanied by structural (corundum to monclinic) and magnetic (paramagnetic to antiferromagnetic) phase transitions, providing a complex interplay of many degrees of freedom \cite{kalcheim2019,frandsen2019,trastoy2020,barazani2023}.
As a foundation of utilizing V$_2$O$_3$-based neuromorphic device applications that are implementing light as an external stimulus, one needs a better understanding of the phase transitions in V$_2$O$_3$.

We have investigated the effect of laser scanning irradiation on a V$_2$O$_3$ thin film during the IMT/MIT by acquiring photomicrographs.
Here we report on laser-induced phase transitions.
While at the IMT, the metal phase is induced, the insulating phase is predominantly induced at the MIT.
The latter is surprising, because one might expect the main effect of the laser irradiation to be heating, which is expected to drive the system towards a more metallic state.

%\section{Material and Methods}
%\section{\label{sec:MatandMeth}}

\begin{figure}
\includegraphics{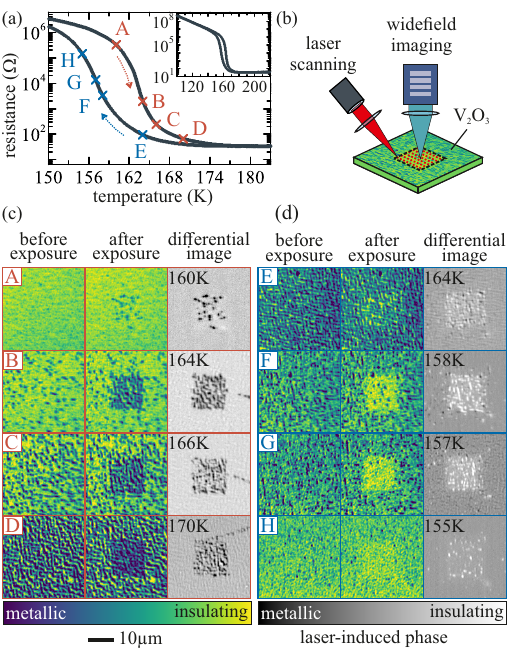}
\caption{
(a) Zoom into the phase transition region of the $R(T)$ curve.
The red and blue dashed arrows indicate the heating and cooling branch, respectively.
Inset: Full $R(T)$ curve of the V$_2$O$_3$ thin film sample under investigation with same units as in the main graph.
The capital letters mark the set-temperatures in one thermal cycle, at which separated squares in the V$_2$O$_3$ thin film were irradiated using the laser scanning mode.
(b) Simplified schematic of the measurement setup.
Photomicrographs are acquired by widefield microscopy before and after laser scanning within the dashed square of pristine areas.
(c) Photomicrograph series demonstrating the laser scanning irradiation effect acquired during one heating run.
For the differential images, the image before laser irradiation was subtracted from the image after scanning.
The image contrasts were optimized for every temperature.
(d) Photomicrograph series acquired during cooling, analogous to (c).
}
\label{fig:B}
\end{figure}

The sample under investigation is a continuous 300-nm-thick rf-sputtered V$_2$O$_3$ film on an r-cut sapphire substrate with gold electrodes evaporated on top (see Ref.~\cite{stewart2012} for details); Fig.~\ref{fig:B}(a) shows its $R(T)$ curve for both heating and cooling. 
The resistance change of four orders of magnitude at the phase transition is a sign of high film quality.

We used the same optical microscopy setup as in Lange \textit{et al.} \cite{lange2021}, where we have demonstrated that the change in reflectivity allows for imaging the separation of the insulating and metallic phases with 0.5\,$\mu$m spatial resolution. 
As described in Refs.~\cite{lange2017, lange2018}, the setup is a cryogenic combined widefield and laser scanning microscope.
A sketch of the experimental setup is shown in Fig.~\ref{fig:B}(b), and a detailed description is given in Ref.~\cite{luibrand-suppl} (Sec.\,\ref{sec:sample:setup}).

For this study, the photomicrographs are acquired in the widefield mode with an monochromatic (532\,nm) LED illumination, and the laser scanning mode is used for exposing the sample under investigation to laser light of 405\,nm wavelength.
The theoretical laser spot diameter on the thin-film surface is 322\,nm.
We have measured the incident laser power on the sample surface by replacing the sample with a photodiode.
The maximum laser power is \,919\,$\mu$W.
However, different laser powers were chosen, throughout this study. 
For exposure, the selected area is scanned line by line (line separation 100\,nm) with the laser in the continuous wave mode as depicted in Fig.~\ref{fig:B}(b). 
During each line scan, the laser spot stays at a position for 13\,ms then it moves to the next position at a distance of 100\,nm.

The procedure to investigate the effect of laser scanning irradiation is the following: 
\begin{enumerate}
    \item A photomicrograph (32$\times$32\,$\mu$m$^2$) of the thin film is acquired.
    \item A selected area (14$\times$14\,$\mu$m$^2$) is exposed to laser light.
    \item A second photomicrograph is acquired. 
    \item For the differential image, the first photomicrograph is subtracted from the second. 
\end{enumerate}
With this procedure, the laser-induced changes from the metallic to the insulating and from the insulating to the metallic phases are mapped.

%%%%%%%%%%%% Fig.1 %%%%%%%%%%%%%%%%%%%%%%%%%%%%%%%%%%%%%%%%%
%%%%%%%%%%%% Fig.1 %%%%%%%%%%%%%%%%%%%%%%%%%%%%%%%%%%%%%%%%%

We have exposed a V$_2$O$_3$ thin film by laser light at different temperatures during the IMT/MIT (Fig.~\ref{fig:B}).
The $R(T)$ curve shows a hysteresis of several Kelvins due to the first-oder nature of the IMT/MIT \cite{mcwhan1973}.
During heating of the sample from 80\,K to room temperature, the heating was interrupted four times during the IMT, and a pristine (unirradiated) 14$\times$14\,$\mu$m area was exposed with 0.1\,mW laser power according to the procedure described above.
The same procedure was chosen for cooling from room temperature to 80\,K with 0.6\,mW laser power.
Note, that we chose different pristine areas on the sample for each scanning procedure.
The photomicrographs in Fig.~\ref{fig:B}(c),(d) (in particular, D and E) show the characteristic herringbone domain pattern due to strain-induced separation of the metallic and insulating phases \cite{mcleod2016, lange2021}. 
A crucial result of this study is that during heating, mostly metallic domains are induced (dark patches in the differential images in Fig.~\ref{fig:B}(c) and (d)), whereas during cooling, predominantly insulating domains are created by laser irradiation (bright patches in the differential images in Fig.~\ref{fig:B}(c) and (d)). 
Furthermore, laser scanning irradiation does not smear or destroy the herringbone domain patterns but modifies them by changing the ratio of the metallic and insulating surface areas.
In an additional experiment on a patterned V$_2$O$_3$ thin film microbridge (see Fig.~\ref{fig:E} in Ref.~\cite{luibrand-suppl}) we show, that indeed the laser-scanning-induced phase change has a significant effect on the film resistivity.
Moreover, we observe that the size of the laser-induced domains is similar to the size of the domains of the temperature-driven phase transition without laser exposure.
This indicates, that domain walls permeate the whole film, and that the laser-induced phase transition is not only due to a surface effect.

We also modified the laser irradiation procedure to determine the power thresholds for laser-induced phase changes during heating and cooling (see Sec.\,\ref{sec:Thresholds} in Ref.~\cite{luibrand-suppl} for details).
It was found that laser irradiation already at relatively low power levels induces a phase change during heating (i.e., IMT), while during cooling (i.e., MIT), the required laser power is considerably larger.
Moreover, the temperature range, in which a laser-induced phase change was observable, is larger during IMT, compared to MIT.
Hence, the heating branch of the transition seems to be more sensitive to laser irradiation than the cooling branch.

%%%%%%%%%%%% Fig.2 %%%%%%%%%%%%%%%%%%%%%%%%%%%%%%%%%%%%%%%%%
\begin{figure}
\includegraphics{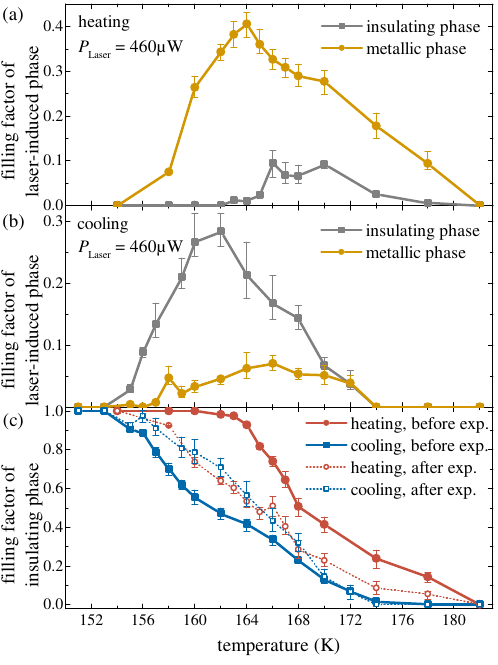}
\caption{
Temperature dependence of metal and insulator filling factors affected by laser irradiation of 460\,$\mu$W.
(a) Filling factors of the metallic and insulating phases induced by laser irradiation during heating.
(b) Same filling factors during cooling.
(c) Filling factors of the insulating phase during one thermal cycle.
}
\label{fig:C}
\end{figure}
%%%%%%%%%%%% Fig.2 %%%%%%%%%%%%%%%%%%%%%%%%%%%%%%%%%%%%%%%%%

For a quantitative analysis of the laser-induced phase changes, we investigated the filling factors, i.e., the ratio between the surface area of a particular phase and the total surface area of interest at a laser power of 460\,$\mu$W. 
This laser power was chosen because it is far from the minimum and maximum threshold values for both cooling and heating (see Sec.\,\ref{sec:Thresholds} in Ref.~\cite{luibrand-suppl}).
As mentioned above, the metallic and insulating thin-film areas can be clearly distinguished by their brightness level in the photomicrographs (see Fig.~\ref{fig:B} and \cite{lange2021}). 
However, these brightness levels change from micrograph to micrograph. 
Therefore, the following evaluation procedure was used for determining the filling factors. 
We visually inspected the (differential) micrographs in Fiji \cite{schindelin2012} and chose an upper and lower limit for the brightness that separates the particular two phases. 
A third brightness level was calculated by averaging the upper and lower limit.
For these three brightness values, we calculated the filling factors, which correspond to the data points and the error bars in Fig.~\ref{fig:C}, using a Python routine.

Figure~\ref{fig:C}(a) and (b) show the temperature dependence of the filling factors of the laser-induced metallic and insulating phases during heating and cooling, respectively.
Except for the fact that in the heating case, predominantly the metallic, and in the cooling case, predominantly the insulating phase is created, the overall behavior is very similar: the laser irradiation switches a considerable percentage of the thin-film area during the IMT from insulating to metallic (40\% at 164\,K) and during the MIT from metallic to insulating (28\% at 162\,K).
Furthermore, in both cases, there is a temperature interval in which a smaller amount of the minor phase (insulating at the IMT and metallic at the MIT) is induced as well.

Figure~\ref{fig:C}(c) shows the filling factor of the insulating phase before and after laser irradiation.
The error bars of the before-irradiation graphs are smaller compared to the after-irradiation graphs because a larger area was used for determining the filling factors before irradiation.
Like the $R(T)$-curves shown in Fig.~\ref{fig:B}(a), the before-irradiation curves are hysteretic due to the first-order character of the IMT/MIT phase transitions.
Most importantly, the after-irradiation curves (dashed lines) lay between the curves before irradiation and are almost identical for temperatures at 168\,K and below (solid lines).
This implies that after irradiation, it is no longer possible to decide (by means of the filling factor) whether the film was in the heating or cooling branch.
In this sense, laser irradiation above the power threshold erases the thermal history of the V$_2$O$_3$ thin film.

%\section{Discussion}
%\section{\label{sec:Diss}}
%
The most intriguing result of this study is that laser irradiation predominantly induces insulating domains in large amounts during cooling.
Considering the laser spot as a local heat source, it is not surprising that laser irradiation induces metallic domains since the metallic phase is the high-temperature phase of the Mott transition.
Indeed, we observe this behavior during heating of the sample through the IMT.
However, the predominant laser-induced creation of insulating domains during cooling is puzzling. 
In other words, why does the effect of laser irradiation depend on the thermal history? 
To answer this question, the thermodynamic nature of the MIT/IMT of the V$_2$O$_3$ thin film must be considered.

The thermodynamic description of V$_2$O$_3$ thin films is the subject of ongoing research \cite{castellani1979, kotliar2000, diaz2023}.
It is recognized that at the MIT, three phase transitions are coupled: first, an electronic transition due to Mott-Hubbard charge carrier localization, second, a magnetic transition from a paramagnetic-to-antiferromagnetic state, and third, a structural transition from corundum to a monoclinic crystal structure. 
Furthermore, as indicated in Ref.~\cite{lange2021}, the characteristic herringbone domain pattern resembles the solutions of the Cahn-Larch\'e equation describing a phase separation due to spinodal decomposition \cite{garcke2005a}.
Other studies also indicate that the IMT/MIT is associated with a spinodal instability \cite{bar2018,kundu2020}.
Despite those intricate details, the $R(T)$-hysteresis and the associated latent heat at the IMT and MIT clearly indicate a first-order phase transition. 
Consequently, it is safe to assume that during heating the V$_2$O$_3$ thin film is in a metastable superheated state, and during cooling it is in a metastable supercooled state.
The most likely interpretation of our results is that the perturbation releases the V$_2$O$_3$ thin film from its metastable state, and when the laser spot moves to a new position, the thin film relaxes locally into a stable state.
The general details of the perturbation induced by the laser beam are elusive and additional effects might be at play, for instance, a photoinduced change of the 3$d$-orbital occupation could trigger the formation of metallic nanodroplets in the IMT \cite{ronchi2019}.
With respect to the MIT we measured electrical transport properties during minor thermal cycles to investigate how heating affects a V$_2$O$_3$ thin film in the critical state.
This procedure emulates a heat source that changes the temperature gradually on a large length scale.
A detailed description and discussion of these additional experiments can be found in Ref.~\cite{luibrand-suppl} (Sec.~\ref{sec:therm:eff}).
Although we found that a minor heating cycle can indeed increase the resistance, the effect is not strong enough to account for what we observed in the laser irradiation experiment.
%

%%%%%%%%%%%% Fig.3 %%%%%%%%%%%%%%%%%%%%%%%%%%%%%%%%%%%%%%%%%
\begin{figure}[b]
\includegraphics{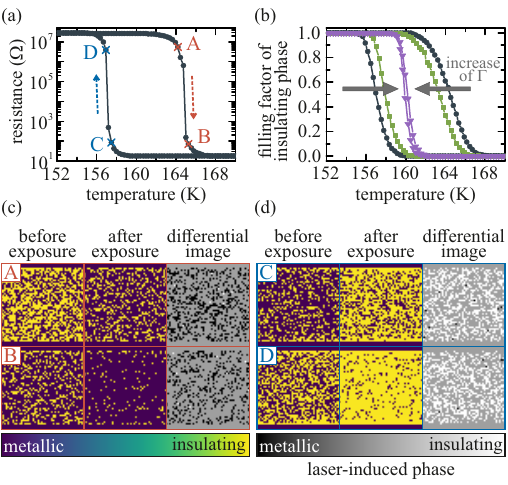}
\caption{
Results of numerical simulations based on a 2D resistor network model.
(a) $R(T)$ curves; arrows indicate heating (red) and cooling (blue) branches.
(b) Filling factors of the insulating phase vs.~$T$, for various values of laser intensity $\Gamma/k_{\rm B} =0$ (grey circles), 600\,K (green squares) and 1200\,K (purple triangles).
(c) Selected network maps before laser exposure, after laser exposure with $\Gamma/k_{\rm B}=600$\,K, and their corresponding differential images for two temperature values in the heating branch, as indicated with capital letters in (a).
(d) same as in (c) -- for the cooling branch.
}
\label{fig:F}
\end{figure}
%%%%%%%%%%%% Fig.3 %%%%%%%%%%%%%%%%%%%%%%%%%%%%%%%%%%%%%%%%%

To elaborate on the matter of metastability, we discuss a phenomenological numerical model.
In a two-dimensional (2D) resistor network, each site can either be metallic or insulating, depending on the local temperature, which is updated at every simulation step.
Solving Kirchhoff's laws, currents and voltages on each site are determined.
A first-order Landau-type free energy functional provides stable and metastable states separated by a barrier.
The laser irradiation is implemented in the model by decreasing the barrier height to drive transitions between insulating and metallic states.
For each site, the escape from a metastable state was determined by comparing random numbers to Boltzmann factors with a constant number of iterations.
If the escape criterion is fulfilled, the transition probability is given by a Boltzmann distribution; see Sec.\,\ref{sec:Mott:res:net:simu} in Ref.~\cite{luibrand-suppl} for details.

Our model reproduces the hysteresis in the $R(T)$ curves, as shown in Fig.~\ref{fig:F}(a). 
Figure~\ref{fig:F}(b) shows the evolution of the filling factor of the insulating phase at different laser intensities.
With increasing laser intensity the hysteresis is diminished, and eventually it is completely suppressed, similarly to the experimental results in Fig.~\ref{fig:C}(c).
In Fig.~\ref{fig:F}(c) we present a simulation series of selected network maps before and after laser exposure with moderate laser intensity ($\Gamma/k_{\rm B}$ = 600\,K), and their corresponding differential images.
For heating, the laser induces predominantly metallic sites (Fig.~\ref{fig:F}(c)), for cooling insulating sites (Fig.~\ref{fig:F}(d)). 
We note that during heating, the majority of sites goes into a metallic state (indicated by black squares in the differential maps), while a few sites revert back to an insulating state (indicated by white squares in the differential maps); this reproduces very well the experimental results.
Moreover, this behavior reverses in the cooling branch (Fig.~\ref{fig:F}(d), again in agreement with experimental observations.
So, according to the simulations, the laser irradiation enhances the process of relaxing the system from a metastable state.

%\section{Conclusion}
%\section{\label{sec:Conc}}
%\sgc{Conclusion is new}

In conclusion, we can state that a V$_2$O$_3$ thin film in the metastable state at the IMT/MIT is highly susceptible to a local perturbation by a focused laser beam at the sub-micrometer scale. 
Whether the laser perturbation predominantly creates a metallic or insulating phase depends on whether the film is in the heating or cooling branch.
Our observations tell that the laser beam brings the system from a supercooled/superheated state to the equilibrium one, provided that the laser power is sufficient to overcome corresponding energy barriers.
This implies, that laser scanning irradiation can be used to effectively erase thermal history and provides a novel method to reset memristive devices locally.
Our experiments may also be considered as a clear indication that external stimuli on a strongly correlated thin film in a critical state can result in unexpected and presumably non-thermal effects.
Non-thermal effects have been investigated in ultra-fast pump-probe experiments on VO$_2$ and V$_2$O$_3$ \cite{giorgianni2019,cavalleri2001,wall2018,ronchi2019,otto2018}.
In contrast, in our experiment we have a quasi-static setting.
The findings of this study may trigger additional experimental and theoretical studies to explore the thermodynamics of phase transitions and local perturbations in this class of materials.

\begin{acknowledgments}
T.L. acknowledges support from the Cusanuswerk, Bisch\"ofliche Studienf\"orderung,
F.T. acknowledges support from the Landesgraduiertenf\"orderung Baden-W\"urttemberg and
M.R. acknowledges support from the French ANR ‘‘MoMA’’ project ANR-19-CE30-0020.
This project has received funding from the European Union's Horizon Europe research and innovation programme under grant agreement No. 2031928 – ‘‘Highly Energy-Efficient Resistive Switching in Defect- and Strain- Engineered Mott Insulators for Neuromorphic Computing Applications’’.
Views and opinions expressed are however those of the authors only and do not necessarily reflect those of the European union or the European research council.
Work at University of California San Diego (UCSD) was supported through an Energy Frontier Research Center program funded by the US Department of Energy (DOE), Office of Science, Basic Energy Sciences, under Grant DE-SC0019273.
\end{acknowledgments}

\let\addcontentsline\oldaddcontentsline
\clearpage

\widetext
\begin{center}
	\noindent\textbf{\large Supplemental material: Laser-induced quenching of metastability at the Mott-insulator to metal transition}
	
	\normalsize
	\vspace{.3cm}
	
	\noindent{T.~Luibrand \textit{et al.}}
	\\
\end{center}
\vspace{0.2cm}

\tableofcontents

\newpage
\setcounter{figure}{0}
\setcounter{equation}{0}
\setcounter{table}{0}
\renewcommand{\thefigure}{S\arabic{figure}}
\renewcommand{\tablename}{Supplementary Table}
\renewcommand{\theequation}{S\arabic{equation}}

\section{Sample and Measurement Setup}
\label{sec:sample:setup}

\begin{figure}
	\includegraphics{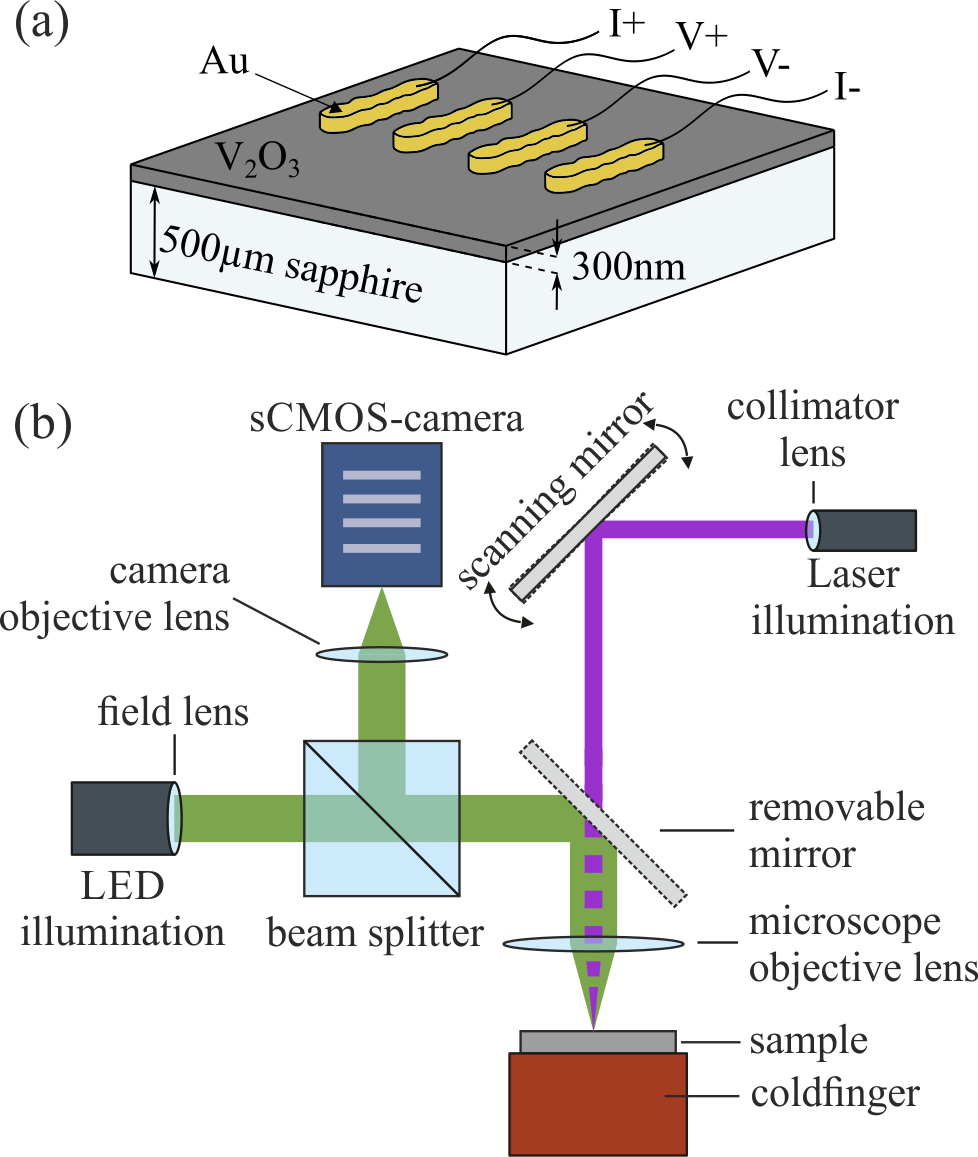}
	\caption{
		(a) V$_2$O$_3$ thin-film sample under investigation.
		Four parallel gold electrodes are used for measuring the $R(T)$ relation.
		(b) Simplified schematic diagram of the combined widefield and laser scanning microscope.
		The removable mirror is used to switch between the two operation modes.}
	\label{fig:A}
\end{figure}

The sample under investigation (Fig.~\ref{fig:A}(a)) is a continuous 300-nm-thick rf-sputtered V$_2$O$_3$ film on an r-cut sapphire substrate.
Using a shadow mask, four 2-mm-long parallel gold electrodes with a distance of 1.5\,mm were evaporated as contacts for resistance measurements.
The sample is mounted in vacuum on the coldfinger of a liquid Helium continuous-flow cryostat with a temperature range of 4.2\,K to 300\,K.
Electric transport measurements were performed using a Keithley 2400 SourceMeter configured as a current source.

The experimental setup is a cryogenic combined widefield and laser scanning microscope, which is described in detail in \cite{lange2017s, lange2018s}.
Figure~\ref{fig:A}(b) shows a simplified schematic of the microscope setup that can be operated either in the widefield or in the laser scanning mode.

The illumination for the widefield microscope is realized by a LED with 532\,nm wavelength and follows a Köhler illumination scheme.
The depicted field lens is used to define the illuminated sample area.
After passing a beam splitter, the light is reflected by a removable mirror and is focused onto the sample by the microscope objective lens.
The microscope objective has a high numerical aperture (NA = 0.8), which allows for a working distance of 1\,mm.
After reflection from the sample, the light travels the same way back, is deflected by the beam splitter and then focused onto the sensor of a low noise sCMOS camera.
This widefield microscope has a field of view (FOV) of 500$\times$500\,$\mu$m$^2$ and a spatial resolution of $0.48\,\mu$m.
The entire optical path of the widefield microscope is polarization sensitive, which however is not relevant for this study.
%

%%%%%%%%% Fig. S1 %%%%%%%%%%%%
%%%%%%%%% Fig. S1 %%%%%%%%%%%%	
%
The illumination for the laser scanning microscope is provided by a 405-nm-laser diode coupled into a single mode fiber.
The laser light is collimated and thereafter reflected by a fast steering mirror, that serves as scanning mirror.
In order to use this operation mode, the removable mirror is not inserted in the beam path.
The laser beam is focused onto the sample by the microscope objective lens.
Assuming a Gaussian beam profile, the theoretical laser spot diameter is 322\,nm (1/$e^2$ diameter).
The FOV of the laser scanning microscope is identical with the FOV of the widefield microscope.
The polarization sensitive components and the detector unit of the laser light are not included here, because in this study, we use the laser scanning microscope exclusively for irradiation.

\section{Microbridge experiment}
\label{sec:Mic:dev:exp}

%%%%%%%%% Fig. S2 %%%%%%%%%%%%
\begin{figure}
	\includegraphics{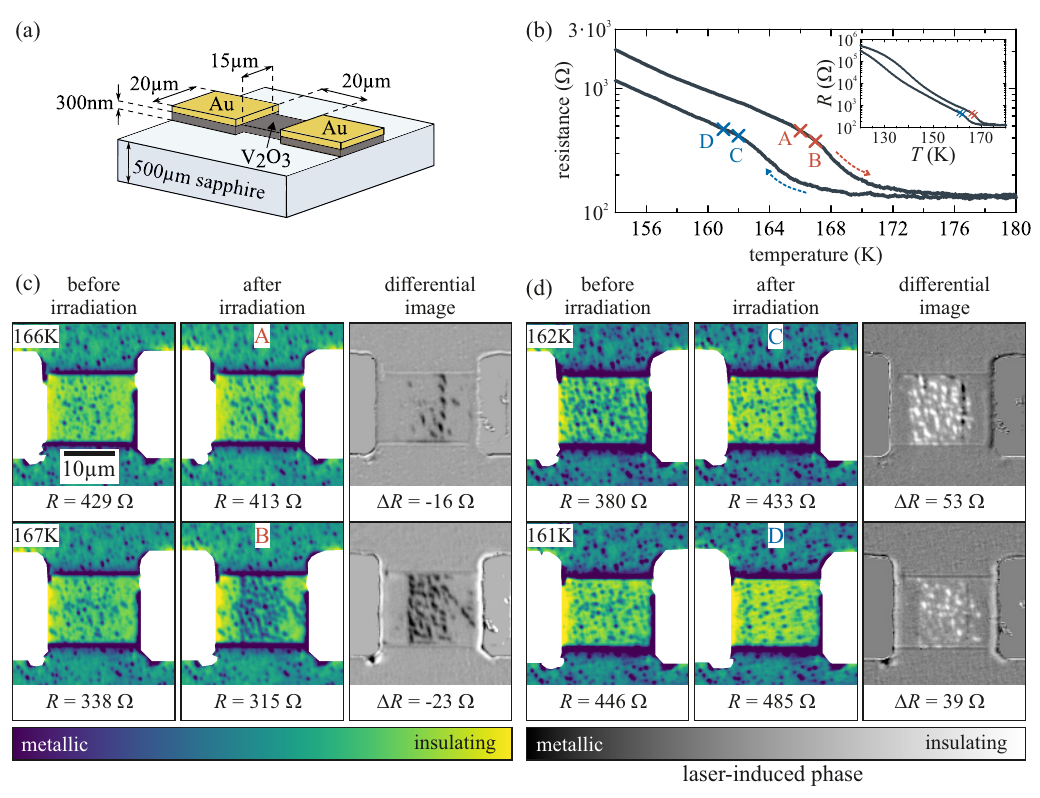}
	\caption{
		Effect of laser irradiation on the resistance of a patterned V$_2$O$_3$ microbridge.
		(a) Schematic of planar microbridge device under test. 
		(b) $R(T)$ curves; red and blue dashed arrows indicate the heating and cooling branch, respectively.
		Capital letters mark the set-temperatures, at which the device was irradiated. 
		Inset: $R(T)$ curves on expanded $T$ scale.
		(c) Photomicrographs demonstrating the laser irradiation effet acquired during one heating run.
		For the differential images, the image before laser irradiation was subtracted from the image after scanning.
		The image contrasts on the microbridge were optimized for both temperatures.
		The two Au electrodes used for the two point measurements are the rectangles in overload at the left and right, respectively.
		The device resistance and the device resistance change due to laser irradiation are shown below the micrographs.  
		(d) analog to (c), but in the cooling branch.
	}
	\label{fig:E}
\end{figure}
%%%%%%%%% Fig. S2 %%%%%%%%%%%%

In order to confirm the hypothesis, that laser scanning indeed induces phase transitions, we investigated its effect on 
the electrical properties of a V$_2$O$_3$ microbridge device in an additional experiment.
The sample under investigation is a planar microbridge device patterned in a 300-nm-thick film V$_2$O$_3$ thin-film (same deposition procedure as in Sec.~\ref{sec:sample:setup}) using Ar ion milling, followed by Au deposition and lift-off to define Au contacts (see Fig.~\ref{fig:E}(a) for device dimensions).
The $R(T)$ curve Fig.~\ref{fig:E}(b) shows irregular behaviour: at the MIT/MIT, the change in resistance is only about one order of magnitude, and the insulating state still shows a relatively low resistance, i.e., the microbridge seems to be shunted.
Possibly, it is shunted at the edges of the microbridge or the V$_2$O$_3$ film was not completely etched down to the substrate during the Ar-ion milling process leaving a thin ion-damaged residual layer.
Nevertheless, the device is still suitable to qualitatively assess the effect of laser irradiation on the film resistivity.

For the two-terminal layout we assume a well-defined homogeneous current distribution in the microbridge.
To bring the sample in a defined state in the heating branch, we used a thermal cycling procedure.
This implies cooling the sample to a temperature, where the hysteresis is closed (in this case $T$\,=\,80\,K) followed by heating to the target temperature.
The same device was used for each laser-irradiation scan, so thermal cycling is required to bring the sample in a defined pristine state for imaging at every single set temperature.
At every set temperature -- after acquiring a widefield micrograph (see left columns of Fig.~\ref{fig:E}(c) and (d)) -- a single stripe of $\sim$ 10\,$\mu$m width across the middle of the bridge was scanned with the laser beam (same scanning parameters as in the main manuscript).
After irradiation a widefield micrograph was acquired (see middle columns of Fig.~\ref{fig:E}(c) and (d)).
The right columns of Fig.~\ref{fig:E}(c) and (d) show differential images, i.e., the image before irradiation was subtracted from the image after irradiation.
Simultaneously, we measured the device resistance before and after laser-irradiation with a bias current of 1\,$\mu$A.

Analogous to the results shown in the main manuscript, metallic and insulating phases are induced by laser irradiation, which show up as black and white patches, respectively, in the differential images.
Clearly, the data obtained on this microbride show, that in the heating branch predominantly the metallic phase is induced, and that comes with a reduction of the device resistance.
In the cooling branch, mostly the insulating phase is induced, which leads to a significant increase of the device resistance.

\section{Thresholds of laser scanning irradiation}
\label{sec:Thresholds}

We modified the laser irradiation procedure to determine the power thresholds for laser-induced phase changes during heating and cooling. 
Instead of scanning the whole $22\times 22\,\mu$m$^2$ area with the same intensity, four $11\times 11\,\mu$m$^2$ quadrants are exposed with different laser powers of 230\,$\mu$W, 460\,$\mu$W, 689\,$\mu$W, and 919\,$\mu$W. 
The two consecutive quadrants, for which the first shows no and the 
next some laser-induced phase change, are chosen by examining the corresponding photomicrographs. 
Following a nested interval procedure, an additional pristine area is exposed to a quadrant pattern, with the minimum and maximum laser power corresponding to the power levels of the two consecutive quadrants chosen before.
Using the photomicrographs of this pattern, the lower and upper limits of the laser power threshold are estimated, and these limits are averaged to calculate the threshold value.

%%%%%%%%% Fig. S3 %%%%%%%%%%%%
\begin{figure}
	\includegraphics{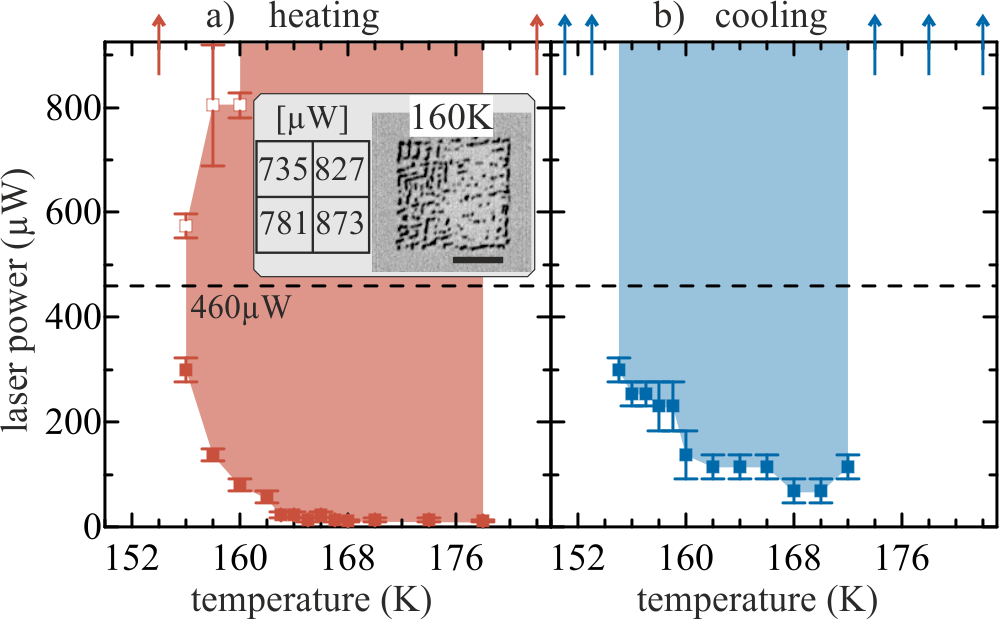}
	\caption{
		Power thresholds vs.~temperature for laser irradiation-induced phase changes.
		The horizontal black dashed line indicates the laser intensity used in Fig.~\ref{fig:C} in the main manuscript.
		The vertical arrows on top of the figure indicate temperatures, at which the laser does not induce a phase change.
		(a) heating branch.
		Inset: Differential image of a laser irradiation pattern at 160\,K.
		The corresponding laser intensities are displayed on the left axis.
		At laser intensities of 827\,$\mu$W and above, a phase change is only induced at the edge of the irradiated area.
		The scale bar indicates 10\,$\mu$m.
		(b) cooling branch.}
	\label{fig:D}
\end{figure}
%%%%%%%%% Fig. S3 %%%%%%%%%%%%

The results are shown in Fig.~\ref{fig:D}.
The temperature range, in which laser irradiation affects the V$_2$O$_3$ film, is larger during heating than during cooling.
Further, the minimal power threshold in the heating branch is 11.5\,$\mu$W (at 168\,K and 178\,K), which corresponds to a  theoretical 0.3\,mW/$\mu$m$^2$ peak intensity assuming a Gaussian point spread function.
However, the minimal power threshold in the cooling branch is six times larger (69\,$\mu$W, peak intensity 1.7\,mW/$\mu$m$^2$).
During heating at low temperatures between 156 K and 160 K, the laser irradiation effect is reduced for high power levels, and the induced phase changes are mainly limited to the edges of the exposed area (see inset Fig.~\ref{fig:D}(a)).
Empty markers in Fig.~\ref{fig:D} indicate the corresponding laser power values.
A similar effect was not observed during cooling.

\section{Thermal effects: minor loop experiment}
\label{sec:therm:eff}

Since the laser irradiation effect discussed in the main manuscript involves focusing the beam onto a very small region, it is important to check whether the observed phenomenology could simply arise from heating.
To address this issue, we performed $R$ vs $T$ measurements on a continuous 100-nm-thick rf-sputtered V$_2$O$_3$ film on r-cut sapphire \cite{trastoy2018a}, with various temperature ramp protocols which would mimic possible laser-induced heating effects.
We are aware of the limitation of the difference between smooth adiabatic thermal cycling of the entire thin film and a local laser-induced thermal perturbation.
Here we focus on measurements conducted on the cooling curve where the experiments on laser irradiation described in the main manuscript showed a surprising increase in the insulating phase fraction in considerably large amounts.
To check whether a simple heating-cooling thermal cycle may induce a similar effect, we performed the following experiments: first, we warmed the sample up to 210\,K to ensure that it is in the fully metallic phase and then slowly cooled it to $T_{\rm reverse}=$154\,K which is in the middle of the resistive transition.
This reversal temperature exhibits the maximum slope of $\log(R)$ vs $T$, where the largest effects are expected in resistance change due to the proximity to the percolation threshold.
After reaching $T_{\rm reverse}$ we performed a minor heating loop, i.e., the sample was heated up by a certain temperature $\Delta T$ and then cooled down to 100\,K.
Several runs were conducted with various values of $\Delta T$.

%%%%%%%%% Fig. S4 %%%%%%%%%%%%
\begin{figure}
	\includegraphics{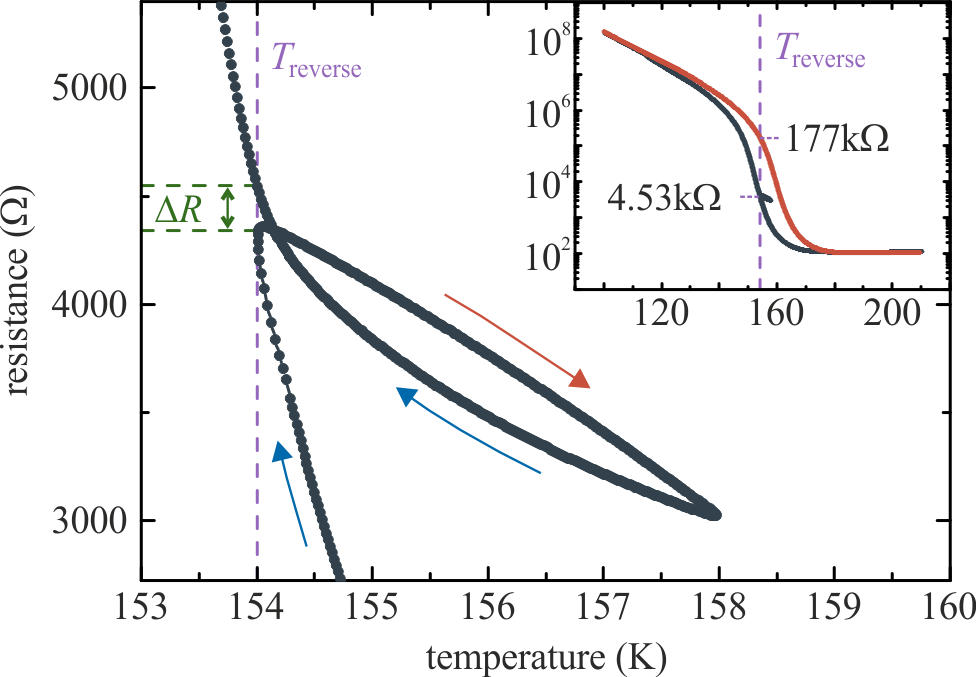}
	\caption{
		Effect of thermal cycle on the cooling branch.
		$R(T)$ curve for 100-nm-thick V$_2$O$_3$ including a minor temperature cycle acquired in the middle of the cooling curve (cooling is marked by blue arrows, heating by the red arrow), starting at $T_{\rm reverse} = $154\,K (as indicated by the purple dashed line).
		The full thermal cycle is shown in the inset (same units as in the main graph), heating curve is indicated in red).
	}
	\label{fig:min:loops}
\end{figure}
%%%%%%%%% Fig. S4 %%%%%%%%%%%%

An example of the resistance recorded during such a procedure is shown in Fig.~\ref{fig:min:loops} for $\Delta T =$ 4\,K which showed the most pronounced effect, i.e., the strongest change $\Delta R$ in resistance at $T_{\rm reverse}$ measured after performing the minor heating loop.
During most of this partial temperature cycle the resistance of the minor heating curve is higher than during cooling, as expected.
However, as $T_{\rm reverse}$ is approached again during cooling the resistance increases and crosses that of the minor heating curve.
After completion of the minor thermal cycle, $R(T_{\rm reverse})=4530\,\Omega$, which corresponds to an increase by $\sim 5\,$\% over the starting resistance $R(T_{\rm reverse})=4330\,\Omega$ measured before initiating the minor temperature cycle.
This increase in resistance is qualitatively similar to the effect observed in the case of laser irradiation, where after exposure the insulating phase fraction has increased (see panels E-H in Fig.~\ref{fig:B}(d) in the main manuscript).
Contrary to the large effect observed in response to laser irradiation, which drives the film close to thermodynamic equilibrium (increase in filling factor of the insulating phase by $\sim 20$\,\% in the middle of the cooling branch; see Fig.~\ref{fig:C}(c) in the main manuscript), the minor heating cycles result in only a slight increase of $\sim 5\,$\% in resistance as compared to the original cooling branch.
We note that the minor thermal cycling loop experiments were repeated for several $\Delta T$ values and that from $\Delta T=2$\,K to 8\,K none showed an increase in resistance at $T_{\rm reverse}$ larger than 5\,\%.
However, quantitatively, it is found that the effect of temperature cycling on the phase fraction is significantly smaller than that of the laser.
Considering Fig.\,\ref{fig:C}(c) in the main manuscript, the filling factors of the insulating phase at 157\,K after laser exposure and at 155\,K before exposure in the cooling branch yield the same value of $\sim 0.9$.
Hence, the impact of the laser irradiation in terms of phase filling factors would correspond to a shift in temperature by $\Delta T_{\rm irr} = -2$\,K.
That, in turn, would correspond to a relative change in resistance of $\Delta R/R = [R(155\rm \,K)-R(157\rm \,K)]/R(155\rm \,K)) \approx 91$\,\% from Fig.~\ref{fig:B}(a) in the main manuscript.
Similar considerations can be made all over the temperature range from 155\,K to 168\,K in Fig.~\ref{fig:C}(c) in the main manuscript, and the relative resistance changes are always well above 5\,\% (see examples in Table~\ref{tab:rel:res:change}).
%
%%%%%% Tab.S1 %%%%%%%%%%%%%%%%%%%%%%%%%%%%%%%%%%%%%%%%%
%\renewcommand{\arraystretch}{1.3}
\begin{table}
	\caption{Estimated relative resistance changes by laser irradiation via filling factors of the insulating phase. }
	\label{tab:rel:res:change}
	\begin{center}
		\tabcolsep5mm
		\begin{tabular}{c c c c c}\hline\hline
			$T_{1}$ 	& $T_{2}$	& $\Delta T_{\rm irr} = T_2-T_1$   	& filling factor of ins. phase 	& $\Delta R/R = \frac{R(T_2)-R(T_1)}{R(T_2)}$     \\
			(K)		& (K)      & (K)	&		& 	(\%)	 		\\\hline
			157		& 155		& $-2$		& $\sim 0.9$	& 	91	\\
			160		& 157		& $-3$		& $\sim 0.8$	& 	96	\\
			162		& 158		& $-4$		& $\sim 0.7$	&	83	\\
			164		& 160		& $-4$		& $\sim 0.55$	&	84	\\
			166		& 164		& $-2$		& $\sim 0.4$	&	35	\\
			168		& 166		& $-2$		& $\sim 0.3$	&	22	\\\hline\hline
		\end{tabular}
	\end{center}
\end{table}
%%%%%% Tab.S1 %%%%%%%%%%%%%%%%%%%%%%%%%%%%%%%%%%%%%%%%%

The conclusion that the effect of minor thermal cycles is considerably smaller than that of the laser is further corroborated by the observation that the phase fractions after laser irradiation from both the heating and cooling curves (Fig.~\ref{fig:C}(c) in the main manuscript) collapse to approximately the same intermediate values between the curves.
If the thermal cycling had a similar effect as the laser irradiation, we would expect it to result in a significant rise in resistance close to the mid-point value between the resistances of the cooling curve (4.33\,k$\Omega$) and heating curve (177\,k$\Omega$) at $T_{\rm reverse}$ (see Fig.~\ref{fig:min:loops}).
Since the resistance only increases to 4530\,$\Omega$ we conclude that the phase fraction after the thermal cycling is much closer to that of the cooling curve than that observed after laser irradiation.

This points to a non-trivial effect of laser irradiation which provides a pathway towards the true equilibrium between the insulating and metallic phases by lowering the free energy barrier between them.

\section{Mott resistor network simulation}
\label{sec:Mott:res:net:simu}

\begin{figure}[b]
	\includegraphics[width=.43\columnwidth]{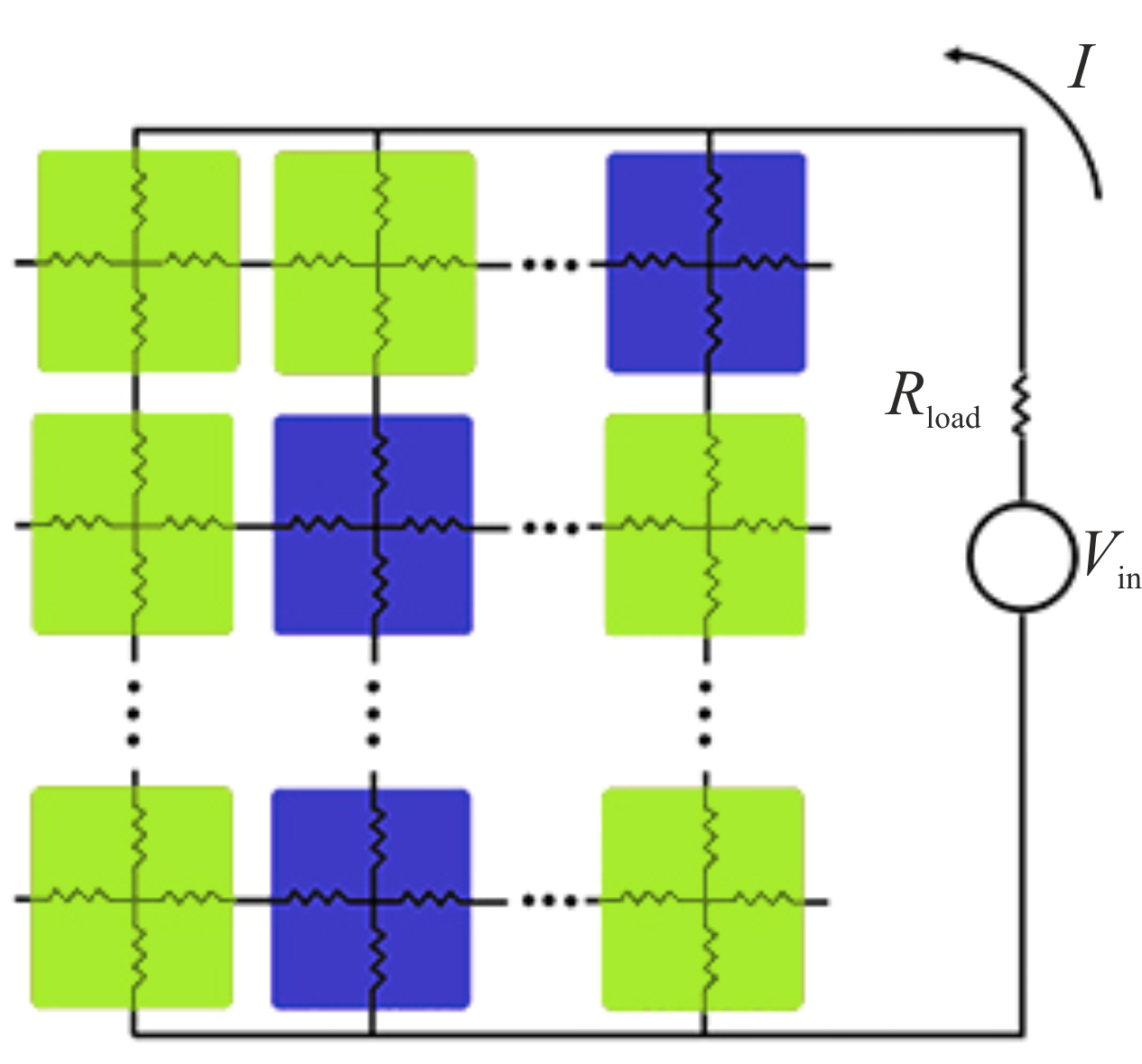}
	\caption{
		Schematic representation of the Mott resistor network model and simulation circuit.
		Light green represents insulating sites with a high resistance $R_{\rm ins}$ and the blue cells represent metallic sites with a low resistance $R_{\rm met}$.
	}
	\label{res_net}
\end{figure}

%\subsection{Model and parameters of the simulations}
%
We model and numerically  solve our system as a Mott resistor network model as schematized in Fig.~\ref{res_net}.
A load resistance $R_{\rm load}$ is making a voltage divider circuit with a resistor tensor network.

%%%%%%%%% Fig. S5 %%%%%%%%%%%%
%%%%%%%%% Fig. S5 %%%%%%%%%%%%

In order to determine the resistance of the entire network, a small bias voltage $V_{\rm in}$ is applied in order to obtain a voltage $V_{\rm S}$ across the system
%
%%%%%%%%% eq.S1 %%%%%%%%%%%%%%%%%%%%%
\begin{equation}
	V_{\rm S}=\frac{R_{\rm S}}{R_{\rm S}+R_{\rm load}}\cdot V_{\rm in}\;,
\end{equation}
%%%%%%%%% eq.S1 %%%%%%%%%%%%%%%%%%%%%
%
where $R_{\rm S}$ denotes the total resistance of the resistor network solved using Kirchhoff laws.
Each site of our  $50\times 50$ grid is represented by four resistors of two possible values, $R_{\rm met} $ and $R_{\rm ins}$.

In order to simulate the first-order transition we assume a free energy  from Landau’s theory with shape \cite{valle2019,stoliar2017}
%
%
%%%%%%%%% eq.S2 %%%%%%%%%%%%%%%%%%%%%
\begin{equation}\label{free_form}
	f(T,\eta)=h(T)\eta+p(T)\eta^2+c\eta^4\;.
	%\label{eq:S2}
\end{equation}
%%%%%%%%% eq.S2 %%%%%%%%%%%%%%%%%%%%%
%
$T$ is the temperature, $\eta$ is an order-parameter-like quantity,
%
%%%%%%%%% eq.S3 %%%%%%%%%%%%%%%%%%%%%
\begin{equation}
	h(T)= h_0 \cdot \frac{T-T_{\rm c}}{T_{\rm c}-T_{\rm A}}\;,
\end{equation}
%%%%%%%%% eq.S3 %%%%%%%%%%%%%%%%%%%%%
%
and
%
%%%%%%%%% eq.S4 %%%%%%%%%%%%%%%%%%%%%
\begin{equation}
	p(T)=p_1 \cdot \frac{T-T_{\rm A}}{T_{\rm A}},
\end{equation}
%%%%%%%%% eq.S4 %%%%%%%%%%%%%%%%%%%%%
%
with a critical temperature $T_{\rm c}$, where the free energy is symmetric, and $T_{\rm A}, \ h_0, \ p_1, \ c$ are constant parameters, specified in Table~\ref{tabella}.

%%%%%% Tab.S1 %%%%%%%%%%%%%%%%%%%%%%%%%%%%%%%%%%%%%%%%%
%\renewcommand{\arraystretch}{1.3}
\begin{table}
	\caption{Constant parameters of the free energy.}
	\label{tabella}
	\begin{center}
		\tabcolsep5mm
		\begin{tabular}{c c c c c}\hline\hline
			$T_{\rm c}$ 	& $T_{\rm A}$	& $h_0/k_{\rm B}$        	& $p_1/k_{\rm B}$ 	& $c/k_{\rm B}$     \\
			(K)		& (K)      & 	(K)		& 	(K)	& (K)		\\\hline
			160		& 87		& -2666.67	& -200	& 4		\\\hline\hline
		\end{tabular}
	\end{center}
\end{table}
%%%%%% Tab.S1 %%%%%%%%%%%%%%%%%%%%%%%%%%%%%%%%%%%%%%%%%

%%%%%%%%% Fig. S6 %%%%%%%%%%%%
\begin{figure}
	\includegraphics{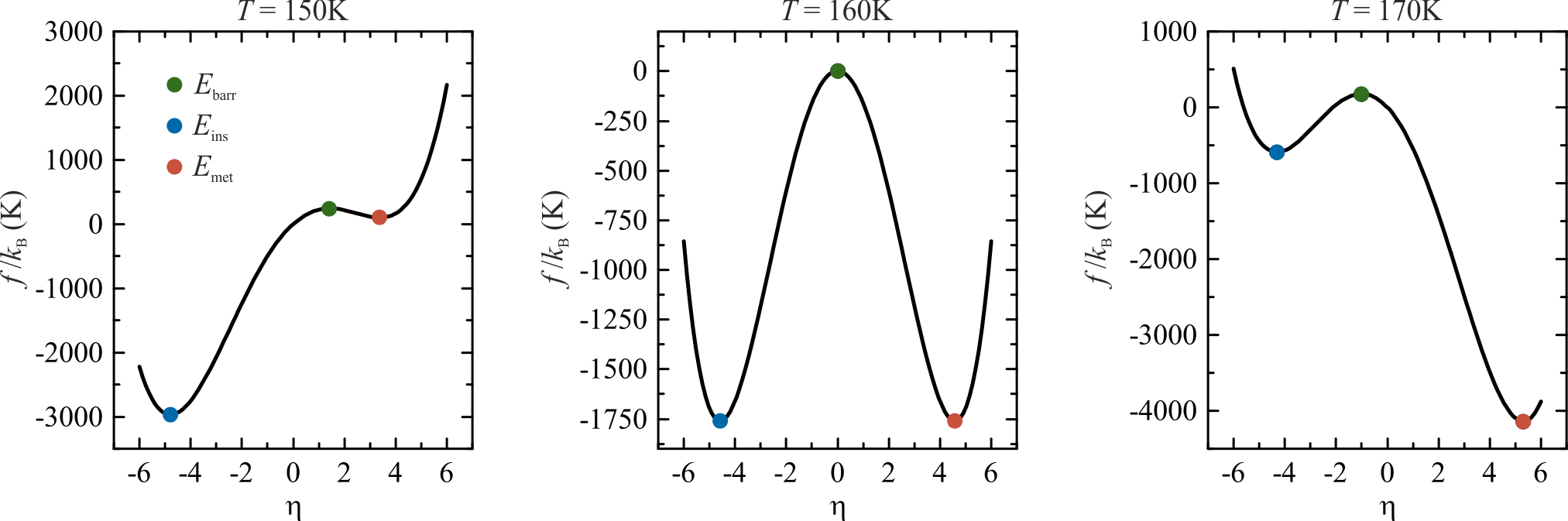}
	\caption{
		Functional form of the free energy $f(\eta)$ of the system with the parameters given in Table.~\ref{tabella} for three selected temperatures: $T=150$\,K $< T_{\rm c}$, $ T=160$\,K $=T_{\rm c}$,  $T=170$\,K $>T_{\rm c}$.
		The green markers show the local maxima $E_{\rm barr}$, the blue and red markers show the minima $E_{\rm ins}$ and $E_{\rm met}$ of the free energy, respectively.
	}
	\label{free_energy}
\end{figure}
%%%%%%%%% Fig. S6 %%%%%%%%%%%%

\begin{figure}
	\includegraphics{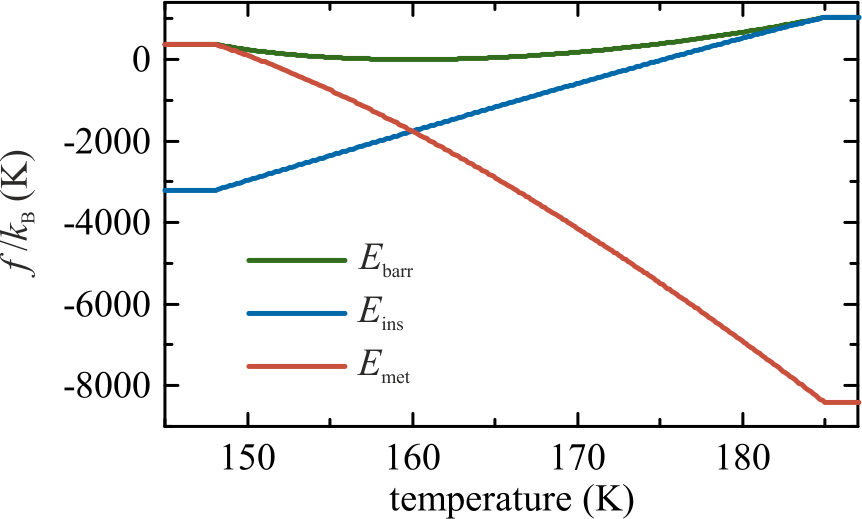}
	\caption{
		Temperature dependence of the energies $E_{\rm met}$ of the metallic state (red), $E_{\rm ins}$ of the insulating state (blue) and $E_{\rm barr}$ of the barrier (green), computed as stationary points of Eq.~(\ref{free_form}).
	}
	\label{ener_VS_T}
\end{figure}

As illustrated for three selected temperatures around $T_{\rm c}$ in Fig.~\ref{free_energy}, the free energy exhibits two minima with energies $E_{\rm met}$ and $E_{\rm ins}$ and one local maximum $E_{\rm barr}$.
The temperature dependence of the energies of these stationary points is shown in Fig.~\ref{ener_VS_T}.

%%%%%%%%% Fig. S7 %%%%%%%%%%%%
%%%%%%%%% Fig. S7 %%%%%%%%%%%%

Each resistor in the network is in a metallic state with energy $E_{\rm met}$ or in an insulating state with energy $E_{\rm ins}$.
In each simulation cycle, we assign an escape criterion to each site, dependent on whether the initial state is metallic or insulating.
\begin{itemize}
	\item The criterion to reach the energy barrier from the metallic state is
	%%%%%%%%%% eq.S5 %%%%%%%%%%%%%%%%%%%%%
	\begin{equation}\label{eq:CM-MI}
		P_{\rm Met\rightarrow  Barrier}=e^{\frac{-(E_{\rm barr}-E_{\rm met}-\Gamma)}{k_{\rm B}T}}\;.
	\end{equation} 
	%%%%%%%%%% eq.S5 %%%%%%%%%%%%%%%%%%%%%
	\item The criterion to reach the barrier from an insulating state is
	%%%%%%%%%% eq.S6 %%%%%%%%%%%%%%%%%%%%%
	\begin{equation}\label{eq:MI-CM}
		P_{\rm Ins\rightarrow  Barrier}=e^{\frac{-(E_{\rm barr}-E_{\rm ins}-\Gamma)}{k_{\rm B}T}}\;.
	\end{equation} 
	%%%%%%%%%% eq.S6 %%%%%%%%%%%%%%%%%%%%%
\end{itemize}
$k_{\rm B}$ denotes the Boltzmann constant and $\Gamma$ the energy contribution of the laser in both equations.
We generate a random number between zero and one and compare it to the escape criterion of Eq.~(\ref{eq:CM-MI}) or Eq.~(\ref{eq:MI-CM}) with a constant number of iterations, respectively.
If the escape criterion exceeds the random number, we assign a resistivity $R_{\rm ins}$, by assuming Boltzmann weights, with probability
%
%%%%%%%%% eq.S7 %%%%%%%%%%%%%%%%%%%%%
\begin{equation}
	\label{eq:prob:ins}
	P_{\rm ins}=\frac{e^{\frac{-E_{\rm ins}}{k_{\rm B}T}}}{e^{\frac{-E_{\rm ins}}{k_{\rm B}T}}+e^{\frac{-E_{\rm met}}{k_{\rm B}T}}},
\end{equation} 
%%%%%%%%% eq.S77 %%%%%%%%%%%%%%%%%%%%%
%
or $R_{\rm met}$ with probability
%
%%%%%%%%% eq.S8 %%%%%%%%%%%%%%%%%%%%%
\begin{equation}
	\label{eq:prob:met}
	P_{\rm met}=\frac{e^{\frac{-E_{\rm met}}{k_{\rm B}T}}}{e^{\frac{-E_{\rm ins}}{k_{\rm B}T}}+e^{\frac{-E_{\rm met}}{k_{\rm B}T}}}\;.
\end{equation} 
%%%%%%%%% eq.S8 %%%%%%%%%%%%%%%%%%%%%
%
$T$ is the local temperature on the site updated at each cycle by the following formula using the heat equation through the discrete Laplacian approximation
%
%%%%%%%%% eq.S9 %%%%%%%%%%%%%%%%%%%%%
\begin{equation}
	T_{ij}=T_{\rm subs}+\frac{V^2}{C_v R_{ij}}+4T_{ij}-k_h \sum_{<ik>}^{\rm first\ neighboors}T_{ik}\;,
\end{equation}
%%%%%%%%% eq.S9 %%%%%%%%%%%%%%%%%%%%%
%
%
%%%%%% Tab.S2 %%%%%%%%%%%%%%%%%%%%%%%%%%%%%%%%%%%%%%%%%
%\renewcommand{\arraystretch}{1.3}
\begin{table}
	\caption{Constant parameters of the simulations.}
	\label{tabella2}
	\begin{center}
		\tabcolsep5mm
		\begin{tabular}{c c c c c }\hline\hline
			$V_{\rm in}$ 		& $C_v$	& $k_h$  & $R_{\rm met}$	& $R_{\rm ins}$     \\
			(V)			&   (J/K)		& 	(J/K)	& 	($\Omega$)		& ($\Omega$) 		\\\hline
			$10^{-5}$	& 1		& 0.166	& 10			& $1.5 \cdot 10^7$ \\\hline\hline
		\end{tabular}
	\end{center}
\end{table}
%%%%%% Tab.S2 %%%%%%%%%%%%%%%%%%%%%%%%%%%%%%%%%%%%%%%%%
%
where the indices $i$ and $j$ denote the coordinates of the site in the network, $T_{\rm subs}$ the substrate temperature, $C_v$ the specific heat of the system, $k_h$ the thermal conductivity and $\sum_{<ik>}^{\rm first\ neighboors}$ denotes the summation over the first neighbouring sites around the one with coordinates $ij$.
The constant parameters used in our calculations are summarized in Table~\ref{tabella2}.

\let\oldaddcontentsline\addcontentsline
\renewcommand{\addcontentsline}[3]{}


\begin{thebibliography}{45}%
	\makeatletter
	\providecommand \@ifxundefined [1]{%
		\@ifx{#1\undefined}
	}%
	\providecommand \@ifnum [1]{%
		\ifnum #1\expandafter \@firstoftwo
		\else \expandafter \@secondoftwo
		\fi
	}%
	\providecommand \@ifx [1]{%
		\ifx #1\expandafter \@firstoftwo
		\else \expandafter \@secondoftwo
		\fi
	}%
	\providecommand \natexlab [1]{#1}%
	\providecommand \enquote  [1]{``#1''}%
	\providecommand \bibnamefont  [1]{#1}%
	\providecommand \bibfnamefont [1]{#1}%
	\providecommand \citenamefont [1]{#1}%
	\providecommand \href@noop [0]{\@secondoftwo}%
	\providecommand \href [0]{\begingroup \@sanitize@url \@href}%
	\providecommand \@href[1]{\@@startlink{#1}\@@href}%
	\providecommand \@@href[1]{\endgroup#1\@@endlink}%
	\providecommand \@sanitize@url [0]{\catcode `\\12\catcode `\$12\catcode
		`\&12\catcode `\#12\catcode `\^12\catcode `\_12\catcode `\%12\relax}%
	\providecommand \@@startlink[1]{}%
	\providecommand \@@endlink[0]{}%
	\providecommand \url  [0]{\begingroup\@sanitize@url \@url }%
	\providecommand \@url [1]{\endgroup\@href {#1}{\urlprefix }}%
	\providecommand \urlprefix  [0]{URL }%
	\providecommand \Eprint [0]{\href }%
	\providecommand \doibase [0]{https://doi.org/}%
	\providecommand \selectlanguage [0]{\@gobble}%
	\providecommand \bibinfo  [0]{\@secondoftwo}%
	\providecommand \bibfield  [0]{\@secondoftwo}%
	\providecommand \translation [1]{[#1]}%
	\providecommand \BibitemOpen [0]{}%
	\providecommand \bibitemStop [0]{}%
	\providecommand \bibitemNoStop [0]{.\EOS\space}%
	\providecommand \EOS [0]{\spacefactor3000\relax}%
	\providecommand \BibitemShut  [1]{\csname bibitem#1\endcsname}%
	\let\auto@bib@innerbib\@empty
	%</preamble>

\subsection*{References}

	\bibitem [{\citenamefont {Schuller}\ \emph {et~al.}(2015)\citenamefont
		{Schuller}, \citenamefont {Stevens}, \citenamefont {Pino},\ and\
		\citenamefont {Pechan}}]{schuller2015}%
	\BibitemOpen
	\bibfield  {author} {\bibinfo {author} {\bibfnamefont {I.~K.}\ \bibnamefont
			{Schuller}}, \bibinfo {author} {\bibfnamefont {R.}~\bibnamefont {Stevens}},
		\bibinfo {author} {\bibfnamefont {R.}~\bibnamefont {Pino}},\ and\ \bibinfo
		{author} {\bibfnamefont {M.}~\bibnamefont {Pechan}},\ }\href
	{https://doi.org/10.2172/1283147} {\bibinfo {title} {Neuromorphic {Computing}
			{\textendash} {From} {Materials} {Research} to {Systems} {Architecture}
			{Roundtable}}},\ \bibinfo {howpublished} {10.2172/1283147} (\bibinfo {year}
	{2015})\BibitemShut {NoStop}%
	\bibitem [{\citenamefont {del Valle}\ \emph {et~al.}(2018)\citenamefont {del
			Valle}, \citenamefont {Ram{\'{\i}}rez}, \citenamefont {Rozenberg},\ and\
		\citenamefont {Schuller}}]{valle2018}%
	\BibitemOpen
	\bibfield  {author} {\bibinfo {author} {\bibfnamefont {J.}~\bibnamefont {del
				Valle}}, \bibinfo {author} {\bibfnamefont {J.~G.}\ \bibnamefont
			{Ram{\'{\i}}rez}}, \bibinfo {author} {\bibfnamefont {M.~J.}\ \bibnamefont
			{Rozenberg}},\ and\ \bibinfo {author} {\bibfnamefont {I.~K.}\ \bibnamefont
			{Schuller}},\ }\bibfield  {title} {\bibinfo {title} {Challenges in materials
			and devices for resistive-switching-based neuromorphic computing},\ }\href
	{https://doi.org/10.1063/1.5047800} {\bibfield  {journal} {\bibinfo
			{journal} {Journal of Applied Physics}\ }\textbf {\bibinfo {volume} {124}},\
		\bibinfo {pages} {211101} (\bibinfo {year} {2018})}\BibitemShut {NoStop}%
	\bibitem [{\citenamefont {Hoffmann}\ \emph {et~al.}(2022)\citenamefont
		{Hoffmann}, \citenamefont {Ramanathan}, \citenamefont {Grollier},
		\citenamefont {Kent}, \citenamefont {Rozenberg}, \citenamefont {Schuller},
		\citenamefont {Shpyrko}, \citenamefont {Dynes}, \citenamefont {Fainman},
		\citenamefont {Frano}, \citenamefont {Fullerton}, \citenamefont {Galli},
		\citenamefont {Lomakin}, \citenamefont {Ong}, \citenamefont {Petford-Long},
		\citenamefont {Schuller}, \citenamefont {Stiles}, \citenamefont {Takamura},\
		and\ \citenamefont {Zhu}}]{hoffmann2022}%
	\BibitemOpen
	\bibfield  {author} {\bibinfo {author} {\bibfnamefont {A.}~\bibnamefont
			{Hoffmann}}, \bibinfo {author} {\bibfnamefont {S.}~\bibnamefont
			{Ramanathan}}, \bibinfo {author} {\bibfnamefont {J.}~\bibnamefont
			{Grollier}}, \bibinfo {author} {\bibfnamefont {A.~D.}\ \bibnamefont {Kent}},
		\bibinfo {author} {\bibfnamefont {M.~J.}\ \bibnamefont {Rozenberg}}, \bibinfo
		{author} {\bibfnamefont {I.~K.}\ \bibnamefont {Schuller}}, \bibinfo {author}
		{\bibfnamefont {O.~G.}\ \bibnamefont {Shpyrko}}, \bibinfo {author}
		{\bibfnamefont {R.~C.}\ \bibnamefont {Dynes}}, \bibinfo {author}
		{\bibfnamefont {Y.}~\bibnamefont {Fainman}}, \bibinfo {author} {\bibfnamefont
			{A.}~\bibnamefont {Frano}}, \bibinfo {author} {\bibfnamefont {E.~E.}\
			\bibnamefont {Fullerton}}, \bibinfo {author} {\bibfnamefont {G.}~\bibnamefont
			{Galli}}, \bibinfo {author} {\bibfnamefont {V.}~\bibnamefont {Lomakin}},
		\bibinfo {author} {\bibfnamefont {S.~P.}\ \bibnamefont {Ong}}, \bibinfo
		{author} {\bibfnamefont {A.~K.}\ \bibnamefont {Petford-Long}}, \bibinfo
		{author} {\bibfnamefont {J.~A.}\ \bibnamefont {Schuller}}, \bibinfo {author}
		{\bibfnamefont {M.~D.}\ \bibnamefont {Stiles}}, \bibinfo {author}
		{\bibfnamefont {Y.}~\bibnamefont {Takamura}},\ and\ \bibinfo {author}
		{\bibfnamefont {Y.}~\bibnamefont {Zhu}},\ }\bibfield  {title} {\bibinfo
		{title} {Quantum materials for energy-efficient neuromorphic computing:
			{O}pportunities and challenges},\ }\href {https://doi.org/10.1063/5.0094205}
	{\bibfield  {journal} {\bibinfo  {journal} {{APL} Materials}\ }\textbf
		{\bibinfo {volume} {10}},\ \bibinfo {pages} {070904} (\bibinfo {year}
		{2022})}\BibitemShut {NoStop}%
	\bibitem [{\citenamefont {Park}\ \emph {et~al.}(2022)\citenamefont {Park},
		\citenamefont {Deng}, \citenamefont {Manna}, \citenamefont {Islam},
		\citenamefont {Yu}, \citenamefont {Yuan}, \citenamefont {Fong}, \citenamefont
		{Chubykin}, \citenamefont {Sengupta}, \citenamefont {Sankaranarayanan},\ and\
		\citenamefont {Ramanathan}}]{park2022}%
	\BibitemOpen
	\bibfield  {author} {\bibinfo {author} {\bibfnamefont {T.~J.}\ \bibnamefont
			{Park}}, \bibinfo {author} {\bibfnamefont {S.}~\bibnamefont {Deng}}, \bibinfo
		{author} {\bibfnamefont {S.}~\bibnamefont {Manna}}, \bibinfo {author}
		{\bibfnamefont {A.~N. M.~N.}\ \bibnamefont {Islam}}, \bibinfo {author}
		{\bibfnamefont {H.}~\bibnamefont {Yu}}, \bibinfo {author} {\bibfnamefont
			{Y.}~\bibnamefont {Yuan}}, \bibinfo {author} {\bibfnamefont {D.~D.}\
			\bibnamefont {Fong}}, \bibinfo {author} {\bibfnamefont {A.~A.}\ \bibnamefont
			{Chubykin}}, \bibinfo {author} {\bibfnamefont {A.}~\bibnamefont {Sengupta}},
		\bibinfo {author} {\bibfnamefont {S.~K. R.~S.}\ \bibnamefont
			{Sankaranarayanan}},\ and\ \bibinfo {author} {\bibfnamefont {S.}~\bibnamefont
			{Ramanathan}},\ }\bibfield  {title} {\bibinfo {title} {Complex {Oxides} for
			{Brain}‐{Inspired} {Computing}: {A} {Review}},\ }\href
	{https://doi.org/10.1002/adma.202203352} {\bibfield  {journal} {\bibinfo
			{journal} {Advanced Materials}\ }\textbf {\bibinfo {volume} {35}},\ \bibinfo
		{pages} {2203352} (\bibinfo {year} {2022})}\BibitemShut {NoStop}%
	\bibitem [{\citenamefont {Morin}(1959)}]{morin1959}%
	\BibitemOpen
	\bibfield  {author} {\bibinfo {author} {\bibfnamefont {F.~J.}\ \bibnamefont
			{Morin}},\ }\bibfield  {title} {\bibinfo {title} {{Oxides} {Which} {Show} a
			{Metal}-to-{Insulator} {Transition} at the {Neel} {Temperature}},\ }\href
	{https://doi.org/10.1103/physrevlett.3.34} {\bibfield  {journal} {\bibinfo
			{journal} {Physical Review Letters}\ }\textbf {\bibinfo {volume} {3}},\
		\bibinfo {pages} {34} (\bibinfo {year} {1959})}\BibitemShut {NoStop}%
	\bibitem [{\citenamefont {McWhan}\ and\ \citenamefont
		{Remeika}(1970)}]{mcwhan1970}%
	\BibitemOpen
	\bibfield  {author} {\bibinfo {author} {\bibfnamefont {D.~B.}\ \bibnamefont
			{McWhan}}\ and\ \bibinfo {author} {\bibfnamefont {J.~P.}\ \bibnamefont
			{Remeika}},\ }\bibfield  {title} {\bibinfo {title} {Metal-{Insulator}
			{Transition} in {(V$_{1-x}$Cr$_x$)$_2$O$_3$}},\ }\href
	{https://doi.org/10.1103/physrevb.2.3734} {\bibfield  {journal} {\bibinfo
			{journal} {Physical Review B}\ }\textbf {\bibinfo {volume} {2}},\ \bibinfo
		{pages} {3734} (\bibinfo {year} {1970})}\BibitemShut {NoStop}%
	\bibitem [{\citenamefont {McWhan}\ \emph {et~al.}(1973)\citenamefont {McWhan},
		\citenamefont {Menth}, \citenamefont {Remeika}, \citenamefont {Brinkman},\
		and\ \citenamefont {Rice}}]{mcwhan1973}%
	\BibitemOpen
	\bibfield  {author} {\bibinfo {author} {\bibfnamefont {D.~B.}\ \bibnamefont
			{McWhan}}, \bibinfo {author} {\bibfnamefont {A.}~\bibnamefont {Menth}},
		\bibinfo {author} {\bibfnamefont {J.~P.}\ \bibnamefont {Remeika}}, \bibinfo
		{author} {\bibfnamefont {W.~F.}\ \bibnamefont {Brinkman}},\ and\ \bibinfo
		{author} {\bibfnamefont {T.~M.}\ \bibnamefont {Rice}},\ }\bibfield  {title}
	{\bibinfo {title} {Metal-insulator {Transitions} in {Pure} and {Doped}
			{V$_2$O$_3$}},\ }\href {https://doi.org/10.1103/physrevb.7.1920} {\bibfield
		{journal} {\bibinfo  {journal} {Physical Review B}\ }\textbf {\bibinfo
			{volume} {7}},\ \bibinfo {pages} {1920} (\bibinfo {year} {1973})}\BibitemShut
	{NoStop}%
	\bibitem [{\citenamefont {Georges}\ \emph {et~al.}(1996)\citenamefont
		{Georges}, \citenamefont {Kotliar}, \citenamefont {Krauth},\ and\
		\citenamefont {Rozenberg}}]{georges1996}%
	\BibitemOpen
	\bibfield  {author} {\bibinfo {author} {\bibfnamefont {A.}~\bibnamefont
			{Georges}}, \bibinfo {author} {\bibfnamefont {G.}~\bibnamefont {Kotliar}},
		\bibinfo {author} {\bibfnamefont {W.}~\bibnamefont {Krauth}},\ and\ \bibinfo
		{author} {\bibfnamefont {M.~J.}\ \bibnamefont {Rozenberg}},\ }\bibfield
	{title} {\bibinfo {title} {Dynamical mean-field theory of strongly correlated
			fermion systems and the limit of infinite dimensions},\ }\href
	{https://doi.org/10.1103/revmodphys.68.13} {\bibfield  {journal} {\bibinfo
			{journal} {Reviews of Modern Physics}\ }\textbf {\bibinfo {volume} {68}},\
		\bibinfo {pages} {13} (\bibinfo {year} {1996})}\BibitemShut {NoStop}%
	\bibitem [{\citenamefont {Limelette}\ \emph {et~al.}(2003)\citenamefont
		{Limelette}, \citenamefont {Georges}, \citenamefont {Jerome}, \citenamefont
		{Wzietek}, \citenamefont {Metcalf},\ and\ \citenamefont
		{Honig}}]{limelette2003}%
	\BibitemOpen
	\bibfield  {author} {\bibinfo {author} {\bibfnamefont {P.}~\bibnamefont
			{Limelette}}, \bibinfo {author} {\bibfnamefont {A.}~\bibnamefont {Georges}},
		\bibinfo {author} {\bibfnamefont {D.}~\bibnamefont {Jerome}}, \bibinfo
		{author} {\bibfnamefont {P.}~\bibnamefont {Wzietek}}, \bibinfo {author}
		{\bibfnamefont {P.}~\bibnamefont {Metcalf}},\ and\ \bibinfo {author}
		{\bibfnamefont {J.~M.}\ \bibnamefont {Honig}},\ }\bibfield  {title} {\bibinfo
		{title} {Universality and {Critical} {Behavior} at the {Mott} {Transition}},\
	}\href {https://doi.org/10.1126/science.1088386} {\bibfield  {journal}
		{\bibinfo  {journal} {Science}\ }\textbf {\bibinfo {volume} {302}},\ \bibinfo
		{pages} {89} (\bibinfo {year} {2003})}\BibitemShut {NoStop}%
	\bibitem [{\citenamefont {Torrance}\ \emph {et~al.}(1992)\citenamefont
		{Torrance}, \citenamefont {Lacorre}, \citenamefont {Nazzal}, \citenamefont
		{Ansaldo},\ and\ \citenamefont {Niedermayer}}]{torrance1992}%
	\BibitemOpen
	\bibfield  {author} {\bibinfo {author} {\bibfnamefont {J.}~\bibnamefont
			{Torrance}}, \bibinfo {author} {\bibfnamefont {P.}~\bibnamefont {Lacorre}},
		\bibinfo {author} {\bibfnamefont {A.}~\bibnamefont {Nazzal}}, \bibinfo
		{author} {\bibfnamefont {E.}~\bibnamefont {Ansaldo}},\ and\ \bibinfo {author}
		{\bibfnamefont {C.}~\bibnamefont {Niedermayer}},\ }\bibfield  {title}
	{\bibinfo {title} {Systematic study of insulator-metal transitions in
			perovskites {RNiO}$_3$ {(R=Pr,Nd,Sm,Eu)} due to closing of charge-transfer
			gap},\ }\href {https://doi.org/10.1103/physrevb.45.8209} {\bibfield
		{journal} {\bibinfo  {journal} {Physical Review B}\ }\textbf {\bibinfo
			{volume} {45}},\ \bibinfo {pages} {8209} (\bibinfo {year}
		{1992})}\BibitemShut {NoStop}%
	\bibitem [{\citenamefont {Catalano}\ \emph {et~al.}(2018)\citenamefont
		{Catalano}, \citenamefont {Gibert}, \citenamefont {Fowlie}, \citenamefont
		{{\'{I}}{\~{n}}iguez}, \citenamefont {Triscone},\ and\ \citenamefont
		{Kreisel}}]{catalano2018}%
	\BibitemOpen
	\bibfield  {author} {\bibinfo {author} {\bibfnamefont {S.}~\bibnamefont
			{Catalano}}, \bibinfo {author} {\bibfnamefont {M.}~\bibnamefont {Gibert}},
		\bibinfo {author} {\bibfnamefont {J.}~\bibnamefont {Fowlie}}, \bibinfo
		{author} {\bibfnamefont {J.}~\bibnamefont {{\'{I}}{\~{n}}iguez}}, \bibinfo
		{author} {\bibfnamefont {J.-M.}\ \bibnamefont {Triscone}},\ and\ \bibinfo
		{author} {\bibfnamefont {J.}~\bibnamefont {Kreisel}},\ }\bibfield  {title}
	{\bibinfo {title} {{Rare}-earth nickelates {$R$NiO$_3$}: thin films and
			heterostructures},\ }\href {https://doi.org/10.1088/1361-6633/aaa37a}
	{\bibfield  {journal} {\bibinfo  {journal} {Reports on Progress in Physics}\
		}\textbf {\bibinfo {volume} {81}},\ \bibinfo {pages} {046501} (\bibinfo
		{year} {2018})}\BibitemShut {NoStop}%
	\bibitem [{\citenamefont {Biermann}\ \emph {et~al.}(2005)\citenamefont
		{Biermann}, \citenamefont {Poteryaev}, \citenamefont {Lichtenstein},\ and\
		\citenamefont {Georges}}]{biermann2005}%
	\BibitemOpen
	\bibfield  {author} {\bibinfo {author} {\bibfnamefont {S.}~\bibnamefont
			{Biermann}}, \bibinfo {author} {\bibfnamefont {A.}~\bibnamefont {Poteryaev}},
		\bibinfo {author} {\bibfnamefont {A.~I.}\ \bibnamefont {Lichtenstein}},\ and\
		\bibinfo {author} {\bibfnamefont {A.}~\bibnamefont {Georges}},\ }\bibfield
	{title} {\bibinfo {title} {{Dynamical} {Singlets} and
			{Correlation}-{Assisted} {Peierls} {Transition} in {VO$_2$}},\ }\href
	{https://doi.org/10.1103/physrevlett.94.026404} {\bibfield  {journal}
		{\bibinfo  {journal} {Physical Review Letters}\ }\textbf {\bibinfo {volume}
			{94}},\ \bibinfo {pages} {026404} (\bibinfo {year} {2005})}\BibitemShut
	{NoStop}%
	\bibitem [{\citenamefont {Imada}\ \emph {et~al.}(1998)\citenamefont {Imada},
		\citenamefont {Fujimori},\ and\ \citenamefont {Tokura}}]{imada1998}%
	\BibitemOpen
	\bibfield  {author} {\bibinfo {author} {\bibfnamefont {M.}~\bibnamefont
			{Imada}}, \bibinfo {author} {\bibfnamefont {A.}~\bibnamefont {Fujimori}},\
		and\ \bibinfo {author} {\bibfnamefont {Y.}~\bibnamefont {Tokura}},\
	}\bibfield  {title} {\bibinfo {title} {Metal-insulator transitions},\ }\href
	{https://doi.org/10.1103/revmodphys.70.1039} {\bibfield  {journal} {\bibinfo
			{journal} {Reviews of Modern Physics}\ }\textbf {\bibinfo {volume} {70}},\
		\bibinfo {pages} {1039} (\bibinfo {year} {1998})}\BibitemShut {NoStop}%
	\bibitem [{\citenamefont {Zhou}\ and\ \citenamefont
		{Ramanathan}(2015)}]{zhou2015}%
	\BibitemOpen
	\bibfield  {author} {\bibinfo {author} {\bibfnamefont {Y.}~\bibnamefont
			{Zhou}}\ and\ \bibinfo {author} {\bibfnamefont {S.}~\bibnamefont
			{Ramanathan}},\ }\bibfield  {title} {\bibinfo {title} {Mott {Memory} and
			{Neuromorphic} {Devices}},\ }\href
	{https://doi.org/10.1109/jproc.2015.2431914} {\bibfield  {journal} {\bibinfo
			{journal} {Proceedings of the {IEEE}}\ }\textbf {\bibinfo {volume} {103}},\
		\bibinfo {pages} {1289} (\bibinfo {year} {2015})}\BibitemShut {NoStop}%
	\bibitem [{\citenamefont {Kim}\ \emph {et~al.}(2014)\citenamefont {Kim},
		\citenamefont {Seo},\ and\ \citenamefont {Lee}}]{kim2014}%
	\BibitemOpen
	\bibfield  {author} {\bibinfo {author} {\bibfnamefont {B.-J.}\ \bibnamefont
			{Kim}}, \bibinfo {author} {\bibfnamefont {G.}~\bibnamefont {Seo}},\ and\
		\bibinfo {author} {\bibfnamefont {Y.~W.}\ \bibnamefont {Lee}},\ }\bibfield
	{title} {\bibinfo {title} {Bidirectional laser triggering of planar device
			based on vanadium dioxide thin film},\ }\href
	{https://doi.org/10.1364/oe.22.009016} {\bibfield  {journal} {\bibinfo
			{journal} {Optics Express}\ }\textbf {\bibinfo {volume} {22}},\ \bibinfo
		{pages} {9016} (\bibinfo {year} {2014})}\BibitemShut {NoStop}%
	\bibitem [{\citenamefont {Kim}\ \emph {et~al.}(2018)\citenamefont {Kim},
		\citenamefont {Jeong}, \citenamefont {Kim},\ and\ \citenamefont
		{Lee}}]{kim2018}%
	\BibitemOpen
	\bibfield  {author} {\bibinfo {author} {\bibfnamefont {J.}~\bibnamefont
			{Kim}}, \bibinfo {author} {\bibfnamefont {S.~J.}\ \bibnamefont {Jeong}},
		\bibinfo {author} {\bibfnamefont {B.-J.}\ \bibnamefont {Kim}},\ and\ \bibinfo
		{author} {\bibfnamefont {Y.~W.}\ \bibnamefont {Lee}},\ }\bibfield  {title}
	{\bibinfo {title} {Laser-triggered current gating based on photothermal
			effect in {VO$_2$} thin-film device using {CO$_2$} laser},\ }\href
	{https://doi.org/10.1007/s00340-018-6936-7} {\bibfield  {journal} {\bibinfo
			{journal} {Applied Physics B}\ }\textbf {\bibinfo {volume} {124}},\ \bibinfo
		{pages} {67} (\bibinfo {year} {2018})}\BibitemShut {NoStop}%
	\bibitem [{\citenamefont {Seo}\ \emph {et~al.}(2012{\natexlab{a}})\citenamefont
		{Seo}, \citenamefont {Kim}, \citenamefont {Wook~Lee},\ and\ \citenamefont
		{Kim}}]{seo2012}%
	\BibitemOpen
	\bibfield  {author} {\bibinfo {author} {\bibfnamefont {G.}~\bibnamefont
			{Seo}}, \bibinfo {author} {\bibfnamefont {B.-J.}\ \bibnamefont {Kim}},
		\bibinfo {author} {\bibfnamefont {Y.}~\bibnamefont {Wook~Lee}},\ and\
		\bibinfo {author} {\bibfnamefont {H.-T.}\ \bibnamefont {Kim}},\ }\bibfield
	{title} {\bibinfo {title} {Photo-{Assisted} {Bistable} {Switching} {Using}
			{{Mott}} {Transition} in {Two}-{Terminal} {{VO$_2$}} {Device}},\ }\href
	{https://doi.org/10.1063/1.3672812} {\bibfield  {journal} {\bibinfo
			{journal} {Applied Physics Letters}\ }\textbf {\bibinfo {volume} {100}},\
		\bibinfo {pages} {011908} (\bibinfo {year} {2012}{\natexlab{a}})}\BibitemShut
	{NoStop}%
	\bibitem [{\citenamefont {Li}\ \emph {et~al.}(2022)\citenamefont {Li},
		\citenamefont {Xie}, \citenamefont {Zhong}, \citenamefont {Zhang},
		\citenamefont {Fu}, \citenamefont {Zhou}, \citenamefont {Li}, \citenamefont
		{Ni}, \citenamefont {Wang}, \citenamefont {Guo}, \citenamefont {He},
		\citenamefont {Wang}, \citenamefont {Yang}, \citenamefont {Jin},\ and\
		\citenamefont {Ge}}]{li2022a}%
	\BibitemOpen
	\bibfield  {author} {\bibinfo {author} {\bibfnamefont {G.}~\bibnamefont
			{Li}}, \bibinfo {author} {\bibfnamefont {D.}~\bibnamefont {Xie}}, \bibinfo
		{author} {\bibfnamefont {H.}~\bibnamefont {Zhong}}, \bibinfo {author}
		{\bibfnamefont {Z.}~\bibnamefont {Zhang}}, \bibinfo {author} {\bibfnamefont
			{X.}~\bibnamefont {Fu}}, \bibinfo {author} {\bibfnamefont {Q.}~\bibnamefont
			{Zhou}}, \bibinfo {author} {\bibfnamefont {Q.}~\bibnamefont {Li}}, \bibinfo
		{author} {\bibfnamefont {H.}~\bibnamefont {Ni}}, \bibinfo {author}
		{\bibfnamefont {J.}~\bibnamefont {Wang}}, \bibinfo {author} {\bibfnamefont
			{E.-j.}\ \bibnamefont {Guo}}, \bibinfo {author} {\bibfnamefont
			{M.}~\bibnamefont {He}}, \bibinfo {author} {\bibfnamefont {C.}~\bibnamefont
			{Wang}}, \bibinfo {author} {\bibfnamefont {G.}~\bibnamefont {Yang}}, \bibinfo
		{author} {\bibfnamefont {K.}~\bibnamefont {Jin}},\ and\ \bibinfo {author}
		{\bibfnamefont {C.}~\bibnamefont {Ge}},\ }\bibfield  {title} {\bibinfo
		{title} {Photo-{Induced} {Non}-{Volatile} {{VO$_2$}} {Phase} {Transition} for
			{Neuromorphic} {Ultraviolet} {Sensors}},\ }\href
	{https://doi.org/10.1038/s41467-022-29456-5} {\bibfield  {journal} {\bibinfo
			{journal} {Nature Communications}\ }\textbf {\bibinfo {volume} {13}},\
		\bibinfo {pages} {1729} (\bibinfo {year} {2022})}\BibitemShut {NoStop}%
	\bibitem [{\citenamefont {Lee}\ \emph {et~al.}(2012)\citenamefont {Lee},
		\citenamefont {Kim}, \citenamefont {Shin},\ and\ \citenamefont
		{Lee}}]{lee2012}%
	\BibitemOpen
	\bibfield  {author} {\bibinfo {author} {\bibfnamefont {Y.-W.}\ \bibnamefont
			{Lee}}, \bibinfo {author} {\bibfnamefont {E.-S.}\ \bibnamefont {Kim}},
		\bibinfo {author} {\bibfnamefont {B.-S.}\ \bibnamefont {Shin}},\ and\
		\bibinfo {author} {\bibfnamefont {S.-M.}\ \bibnamefont {Lee}},\ }\bibfield
	{title} {\bibinfo {title} {{High}-{Performance} {Optical} {Gating} in
			{Junction} {Device} based on {Vanadium} {Dioxide} {Thin} {Film} {Grown} by
			{Sol}-{Gel} {Method}},\ }\href {https://doi.org/10.5370/jeet.2012.7.5.784}
	{\bibfield  {journal} {\bibinfo  {journal} {Journal of Electrical Engineering
				and Technology}\ }\textbf {\bibinfo {volume} {7}},\ \bibinfo {pages} {784}
		(\bibinfo {year} {2012})}\BibitemShut {NoStop}%
	\bibitem [{\citenamefont {Seo}\ \emph {et~al.}(2012{\natexlab{b}})\citenamefont
		{Seo}, \citenamefont {Kim}, \citenamefont {Kim},\ and\ \citenamefont
		{Lee}}]{seo2012a}%
	\BibitemOpen
	\bibfield  {author} {\bibinfo {author} {\bibfnamefont {G.}~\bibnamefont
			{Seo}}, \bibinfo {author} {\bibfnamefont {B.-J.}\ \bibnamefont {Kim}},
		\bibinfo {author} {\bibfnamefont {H.-T.}\ \bibnamefont {Kim}},\ and\ \bibinfo
		{author} {\bibfnamefont {Y.~W.}\ \bibnamefont {Lee}},\ }\bibfield  {title}
	{\bibinfo {title} {Photo-{{Assisted Electrical Oscillation}} in
			{{Two-Terminal Device Based}} on {{Vanadium Dioxide Thin Film}}},\ }\href
	{https://doi.org/10.1109/JLT.2012.2199466} {\bibfield  {journal} {\bibinfo
			{journal} {Journal of Lightwave Technology}\ }\textbf {\bibinfo {volume}
			{30}},\ \bibinfo {pages} {2718} (\bibinfo {year}
		{2012}{\natexlab{b}})}\BibitemShut {NoStop}%
	\bibitem [{\citenamefont {Driscoll}\ \emph {et~al.}(2009)\citenamefont
		{Driscoll}, \citenamefont {Kim}, \citenamefont {Chae}, \citenamefont
		{Ventra},\ and\ \citenamefont {Basov}}]{driscoll2009}%
	\BibitemOpen
	\bibfield  {author} {\bibinfo {author} {\bibfnamefont {T.}~\bibnamefont
			{Driscoll}}, \bibinfo {author} {\bibfnamefont {H.-T.}\ \bibnamefont {Kim}},
		\bibinfo {author} {\bibfnamefont {B.-G.}\ \bibnamefont {Chae}}, \bibinfo
		{author} {\bibfnamefont {M.~D.}\ \bibnamefont {Ventra}},\ and\ \bibinfo
		{author} {\bibfnamefont {D.~N.}\ \bibnamefont {Basov}},\ }\bibfield  {title}
	{\bibinfo {title} {Phase-transition driven memristive system},\ }\href
	{https://doi.org/10.1063/1.3187531} {\bibfield  {journal} {\bibinfo
			{journal} {Applied Physics Letters}\ }\textbf {\bibinfo {volume} {95}},\
		\bibinfo {pages} {043503} (\bibinfo {year} {2009})}\BibitemShut {NoStop}%
	\bibitem [{\citenamefont {Rana}\ \emph {et~al.}(2020)\citenamefont {Rana},
		\citenamefont {Li}, \citenamefont {Koster},\ and\ \citenamefont
		{Hilgenkamp}}]{rana2020}%
	\BibitemOpen
	\bibfield  {author} {\bibinfo {author} {\bibfnamefont {A.}~\bibnamefont
			{Rana}}, \bibinfo {author} {\bibfnamefont {C.}~\bibnamefont {Li}}, \bibinfo
		{author} {\bibfnamefont {G.}~\bibnamefont {Koster}},\ and\ \bibinfo {author}
		{\bibfnamefont {H.}~\bibnamefont {Hilgenkamp}},\ }\bibfield  {title}
	{\bibinfo {title} {Resistive switching studies in {VO$_2$} thin films},\
	}\href {https://doi.org/10.1038/s41598-020-60373-z} {\bibfield  {journal}
		{\bibinfo  {journal} {Scientific Reports}\ }\textbf {\bibinfo {volume}
			{10}},\ \bibinfo {pages} {3293} (\bibinfo {year} {2020})}\BibitemShut
	{NoStop}%
	\bibitem [{\citenamefont {Gao}\ \emph {et~al.}(2022)\citenamefont {Gao},
		\citenamefont {Ros{\'{a}}rio},\ and\ \citenamefont {Hilgenkamp}}]{gao2022}%
	\BibitemOpen
	\bibfield  {author} {\bibinfo {author} {\bibfnamefont {X.}~\bibnamefont
			{Gao}}, \bibinfo {author} {\bibfnamefont {C.~M.~M.}\ \bibnamefont
			{Ros{\'{a}}rio}},\ and\ \bibinfo {author} {\bibfnamefont {H.}~\bibnamefont
			{Hilgenkamp}},\ }\bibfield  {title} {\bibinfo {title} {Multi-level operation
			in {VO$_2$}-based resistive switching devices},\ }\href
	{https://doi.org/10.1063/5.0077160} {\bibfield  {journal} {\bibinfo
			{journal} {{AIP} Advances}\ }\textbf {\bibinfo {volume} {12}},\ \bibinfo
		{pages} {015218} (\bibinfo {year} {2022})}\BibitemShut {NoStop}%
	\bibitem [{\citenamefont {Kalcheim}\ \emph {et~al.}(2019)\citenamefont
		{Kalcheim}, \citenamefont {Butakov}, \citenamefont {Vargas}, \citenamefont
		{Lee}, \citenamefont {del Valle}, \citenamefont {Trastoy}, \citenamefont
		{Salev}, \citenamefont {Schuller},\ and\ \citenamefont
		{Schuller}}]{kalcheim2019}%
	\BibitemOpen
	\bibfield  {author} {\bibinfo {author} {\bibfnamefont {Y.}~\bibnamefont
			{Kalcheim}}, \bibinfo {author} {\bibfnamefont {N.}~\bibnamefont {Butakov}},
		\bibinfo {author} {\bibfnamefont {N.~M.}\ \bibnamefont {Vargas}}, \bibinfo
		{author} {\bibfnamefont {M.-H.}\ \bibnamefont {Lee}}, \bibinfo {author}
		{\bibfnamefont {J.}~\bibnamefont {del Valle}}, \bibinfo {author}
		{\bibfnamefont {J.}~\bibnamefont {Trastoy}}, \bibinfo {author} {\bibfnamefont
			{P.}~\bibnamefont {Salev}}, \bibinfo {author} {\bibfnamefont
			{J.}~\bibnamefont {Schuller}},\ and\ \bibinfo {author} {\bibfnamefont
			{I.~K.}\ \bibnamefont {Schuller}},\ }\bibfield  {title} {\bibinfo {title}
		{Robust {Coupling} between {Structural} and {Electronic} {Transitions} in a
			{Mott} {Material}},\ }\href {https://doi.org/10.1103/physrevlett.122.057601}
	{\bibfield  {journal} {\bibinfo  {journal} {Physical Review Letters}\
		}\textbf {\bibinfo {volume} {122}},\ \bibinfo {pages} {057601} (\bibinfo
		{year} {2019})}\BibitemShut {NoStop}%
	\bibitem [{\citenamefont {Frandsen}\ \emph {et~al.}(2019)\citenamefont
		{Frandsen}, \citenamefont {Kalcheim}, \citenamefont {Valmianski},
		\citenamefont {McLeod}, \citenamefont {Guguchia}, \citenamefont {Cheung},
		\citenamefont {Hallas}, \citenamefont {Wilson}, \citenamefont {Cai},
		\citenamefont {Luke}, \citenamefont {Salman}, \citenamefont {Suter},
		\citenamefont {Prokscha}, \citenamefont {Murakami}, \citenamefont {Kageyama},
		\citenamefont {Basov}, \citenamefont {Schuller},\ and\ \citenamefont
		{Uemura}}]{frandsen2019}%
	\BibitemOpen
	\bibfield  {author} {\bibinfo {author} {\bibfnamefont {B.~A.}\ \bibnamefont
			{Frandsen}}, \bibinfo {author} {\bibfnamefont {Y.}~\bibnamefont {Kalcheim}},
		\bibinfo {author} {\bibfnamefont {I.}~\bibnamefont {Valmianski}}, \bibinfo
		{author} {\bibfnamefont {A.~S.}\ \bibnamefont {McLeod}}, \bibinfo {author}
		{\bibfnamefont {Z.}~\bibnamefont {Guguchia}}, \bibinfo {author}
		{\bibfnamefont {S.~C.}\ \bibnamefont {Cheung}}, \bibinfo {author}
		{\bibfnamefont {A.~M.}\ \bibnamefont {Hallas}}, \bibinfo {author}
		{\bibfnamefont {M.~N.}\ \bibnamefont {Wilson}}, \bibinfo {author}
		{\bibfnamefont {Y.}~\bibnamefont {Cai}}, \bibinfo {author} {\bibfnamefont
			{G.~M.}\ \bibnamefont {Luke}}, \bibinfo {author} {\bibfnamefont
			{Z.}~\bibnamefont {Salman}}, \bibinfo {author} {\bibfnamefont
			{A.}~\bibnamefont {Suter}}, \bibinfo {author} {\bibfnamefont
			{T.}~\bibnamefont {Prokscha}}, \bibinfo {author} {\bibfnamefont
			{T.}~\bibnamefont {Murakami}}, \bibinfo {author} {\bibfnamefont
			{H.}~\bibnamefont {Kageyama}}, \bibinfo {author} {\bibfnamefont {D.~N.}\
			\bibnamefont {Basov}}, \bibinfo {author} {\bibfnamefont {I.~K.}\ \bibnamefont
			{Schuller}},\ and\ \bibinfo {author} {\bibfnamefont {Y.~J.}\ \bibnamefont
			{Uemura}},\ }\bibfield  {title} {\bibinfo {title} {Intertwined magnetic,
			structural, and electronic transitions in {V$_2$O$_3$}},\ }\href
	{https://doi.org/10.1103/physrevb.100.235136} {\bibfield  {journal} {\bibinfo
			{journal} {Physical Review B}\ }\textbf {\bibinfo {volume} {100}},\ \bibinfo
		{pages} {235136} (\bibinfo {year} {2019})}\BibitemShut {NoStop}%
	\bibitem [{\citenamefont {Trastoy}\ \emph {et~al.}(2020)\citenamefont
		{Trastoy}, \citenamefont {Camjayi}, \citenamefont {del Valle}, \citenamefont
		{Kalcheim}, \citenamefont {Crocombette}, \citenamefont {Gilbert},
		\citenamefont {Borchers}, \citenamefont {Villegas}, \citenamefont
		{Ravelosona}, \citenamefont {Rozenberg},\ and\ \citenamefont
		{Schuller}}]{trastoy2020}%
	\BibitemOpen
	\bibfield  {author} {\bibinfo {author} {\bibfnamefont {J.}~\bibnamefont
			{Trastoy}}, \bibinfo {author} {\bibfnamefont {A.}~\bibnamefont {Camjayi}},
		\bibinfo {author} {\bibfnamefont {J.}~\bibnamefont {del Valle}}, \bibinfo
		{author} {\bibfnamefont {Y.}~\bibnamefont {Kalcheim}}, \bibinfo {author}
		{\bibfnamefont {J.-P.}\ \bibnamefont {Crocombette}}, \bibinfo {author}
		{\bibfnamefont {D.~A.}\ \bibnamefont {Gilbert}}, \bibinfo {author}
		{\bibfnamefont {J.~A.}\ \bibnamefont {Borchers}}, \bibinfo {author}
		{\bibfnamefont {J.~E.}\ \bibnamefont {Villegas}}, \bibinfo {author}
		{\bibfnamefont {D.}~\bibnamefont {Ravelosona}}, \bibinfo {author}
		{\bibfnamefont {M.~J.}\ \bibnamefont {Rozenberg}},\ and\ \bibinfo {author}
		{\bibfnamefont {I.~K.}\ \bibnamefont {Schuller}},\ }\bibfield  {title}
	{\bibinfo {title} {Magnetic field frustration of the metal-insulator
			transition in {V$_2$O$_3$}},\ }\href
	{https://doi.org/10.1103/physrevb.101.245109} {\bibfield  {journal} {\bibinfo
			{journal} {Physical Review B}\ }\textbf {\bibinfo {volume} {101}},\ \bibinfo
		{pages} {245109} (\bibinfo {year} {2020})}\BibitemShut {NoStop}%
	\bibitem [{\citenamefont {Barazani}\ \emph {et~al.}(2023)\citenamefont
		{Barazani}, \citenamefont {Das}, \citenamefont {Huang}, \citenamefont
		{Rakshit}, \citenamefont {Saguy}, \citenamefont {Salev}, \citenamefont {del
			Valle}, \citenamefont {Toroker}, \citenamefont {Schuller},\ and\
		\citenamefont {Kalcheim}}]{barazani2023}%
	\BibitemOpen
	\bibfield  {author} {\bibinfo {author} {\bibfnamefont {E.}~\bibnamefont
			{Barazani}}, \bibinfo {author} {\bibfnamefont {D.}~\bibnamefont {Das}},
		\bibinfo {author} {\bibfnamefont {C.}~\bibnamefont {Huang}}, \bibinfo
		{author} {\bibfnamefont {A.}~\bibnamefont {Rakshit}}, \bibinfo {author}
		{\bibfnamefont {C.}~\bibnamefont {Saguy}}, \bibinfo {author} {\bibfnamefont
			{P.}~\bibnamefont {Salev}}, \bibinfo {author} {\bibfnamefont
			{J.}~\bibnamefont {del Valle}}, \bibinfo {author} {\bibfnamefont {M.~C.}\
			\bibnamefont {Toroker}}, \bibinfo {author} {\bibfnamefont {I.~K.}\
			\bibnamefont {Schuller}},\ and\ \bibinfo {author} {\bibfnamefont
			{Y.}~\bibnamefont {Kalcheim}},\ }\bibfield  {title} {\bibinfo {title}
		{Positive and {Negative} {Pressure} {Regimes} in {Anisotropically} {Strained}
			{V$_2$O$_3$} {Films}},\ }\href {https://doi.org/10.1002/adfm.202211801}
	{\bibfield  {journal} {\bibinfo  {journal} {Advanced Functional Materials}\
		}\textbf {\bibinfo {volume} {33}},\ \bibinfo {pages} {2211801} (\bibinfo
		{year} {2023})}\BibitemShut {NoStop}%
	\bibitem [{\citenamefont {Stewart}\ \emph {et~al.}(2012)\citenamefont
		{Stewart}, \citenamefont {Brownstead}, \citenamefont {Wang}, \citenamefont
		{West}, \citenamefont {Ram{\'i}rez}, \citenamefont {Qazilbash}, \citenamefont
		{Perkins}, \citenamefont {Schuller},\ and\ \citenamefont
		{Basov}}]{stewart2012}%
	\BibitemOpen
	\bibfield  {author} {\bibinfo {author} {\bibfnamefont {M.~K.}\ \bibnamefont
			{Stewart}}, \bibinfo {author} {\bibfnamefont {D.}~\bibnamefont {Brownstead}},
		\bibinfo {author} {\bibfnamefont {S.}~\bibnamefont {Wang}}, \bibinfo {author}
		{\bibfnamefont {K.~G.}\ \bibnamefont {West}}, \bibinfo {author}
		{\bibfnamefont {J.~G.}\ \bibnamefont {Ram{\'i}rez}}, \bibinfo {author}
		{\bibfnamefont {M.~M.}\ \bibnamefont {Qazilbash}}, \bibinfo {author}
		{\bibfnamefont {N.~B.}\ \bibnamefont {Perkins}}, \bibinfo {author}
		{\bibfnamefont {I.~K.}\ \bibnamefont {Schuller}},\ and\ \bibinfo {author}
		{\bibfnamefont {D.~N.}\ \bibnamefont {Basov}},\ }\bibfield  {title} {\bibinfo
		{title} {Insulator-to-{Metal} {Transition} and {Correlated Metallic State} of
			{{V$_2$O$_3$}} {Investigated} by {Optical Spectroscopy}},\ }\href
	{https://doi.org/10.1103/physrevb.85.205113} {\bibfield  {journal} {\bibinfo
			{journal} {Physical Review B}\ }\textbf {\bibinfo {volume} {85}},\ \bibinfo
		{pages} {205113} (\bibinfo {year} {2012})}\BibitemShut {NoStop}%
	\bibitem [{\citenamefont {Lange}\ \emph {et~al.}(2021)\citenamefont {Lange},
		\citenamefont {Gu{\'e}non}, \citenamefont {Kalcheim}, \citenamefont
		{Luibrand}, \citenamefont {Vargas}, \citenamefont {Schwebius}, \citenamefont
		{Kleiner}, \citenamefont {Schuller},\ and\ \citenamefont
		{Koelle}}]{lange2021}%
	\BibitemOpen
	\bibfield  {author} {\bibinfo {author} {\bibfnamefont {M.}~\bibnamefont
			{Lange}}, \bibinfo {author} {\bibfnamefont {S.}~\bibnamefont {Gu{\'e}non}},
		\bibinfo {author} {\bibfnamefont {Y.}~\bibnamefont {Kalcheim}}, \bibinfo
		{author} {\bibfnamefont {T.}~\bibnamefont {Luibrand}}, \bibinfo {author}
		{\bibfnamefont {N.~M.}\ \bibnamefont {Vargas}}, \bibinfo {author}
		{\bibfnamefont {D.}~\bibnamefont {Schwebius}}, \bibinfo {author}
		{\bibfnamefont {R.}~\bibnamefont {Kleiner}}, \bibinfo {author} {\bibfnamefont
			{I.~K.}\ \bibnamefont {Schuller}},\ and\ \bibinfo {author} {\bibfnamefont
			{D.}~\bibnamefont {Koelle}},\ }\bibfield  {title} {\bibinfo {title} {Imaging
			of {{Electrothermal Filament Formation}} in a {{Mott Insulator}}},\ }\href
	{https://doi.org/10.1103/physrevapplied.16.054027} {\bibfield  {journal}
		{\bibinfo  {journal} {Physical Review Applied}\ }\textbf {\bibinfo {volume}
			{16}},\ \bibinfo {pages} {054027} (\bibinfo {year} {2021})}\BibitemShut
	{NoStop}%
	\bibitem [{\citenamefont {Lange}\ \emph {et~al.}(2017)\citenamefont {Lange},
		\citenamefont {Gu{\'e}non}, \citenamefont {Lever}, \citenamefont {Kleiner},\
		and\ \citenamefont {Koelle}}]{lange2017}%
	\BibitemOpen
	\bibfield  {author} {\bibinfo {author} {\bibfnamefont {M.~M.}\ \bibnamefont
			{Lange}}, \bibinfo {author} {\bibfnamefont {S.}~\bibnamefont {Gu{\'e}non}},
		\bibinfo {author} {\bibfnamefont {F.}~\bibnamefont {Lever}}, \bibinfo
		{author} {\bibfnamefont {R.}~\bibnamefont {Kleiner}},\ and\ \bibinfo {author}
		{\bibfnamefont {D.}~\bibnamefont {Koelle}},\ }\bibfield  {title} {\bibinfo
		{title} {A {High}-{Resolution} {Combined} {Scanning} {Laser} and {Widefield}
			{Polarizing} {Microscope} for {Imaging} at {Temperatures} from 4 {{K}} to 300
			{{K}}},\ }\href {https://doi.org/10.1063/1.5009529} {\bibfield  {journal}
		{\bibinfo  {journal} {Review of Scientific Instruments}\ }\textbf {\bibinfo
			{volume} {88}},\ \bibinfo {pages} {123705} (\bibinfo {year}
		{2017})}\BibitemShut {NoStop}%
	\bibitem [{\citenamefont {Lange}(2018)}]{lange2018}%
	\BibitemOpen
	\bibfield  {author} {\bibinfo {author} {\bibfnamefont {M.~M.}\ \bibnamefont
			{Lange}},\ }\href@noop {} {\bibinfo {title} {A {High-Resolution Polarizing
				Microscope} for {Cryogenic} {Imaging}: {{Development}} and {{Application}} to
			{{Investigations}} on {{Twin Walls}} in {{SrTiO$_3$}} and the
			{{Metal-Insulator Transition}} in {{V$_2$O$_3$}}}},\ \bibinfo {howpublished}
	{Ph.D. thesis, Universität Tübingen} (\bibinfo {year} {2018})\BibitemShut
	{NoStop}%
	\bibitem [{lui()}]{luibrand-suppl}%
	\BibitemOpen
	\href@noop {} {}\bibinfo {note} {See {S}upplemental {M}aterial at [{URL}] for
		details on the sample and measurement setup, a microbridge experiment,
		thresholds of laser scanning irradiation, analysis of thermal effects via
		minor loop measurements, and {M}ott resistor network simulation.}\BibitemShut
	{Stop}%
	\bibitem [{\citenamefont {McLeod}\ \emph {et~al.}(2016)\citenamefont {McLeod},
		\citenamefont {van Heumen}, \citenamefont {Ram{\'i}rez}, \citenamefont
		{Wang}, \citenamefont {Saerbeck}, \citenamefont {Gu{\'e}non}, \citenamefont
		{Goldflam}, \citenamefont {Anderegg}, \citenamefont {Kelly}, \citenamefont
		{Mueller}, \citenamefont {Liu}, \citenamefont {Schuller},\ and\ \citenamefont
		{Basov}}]{mcleod2016}%
	\BibitemOpen
	\bibfield  {author} {\bibinfo {author} {\bibfnamefont {A.~S.}\ \bibnamefont
			{McLeod}}, \bibinfo {author} {\bibfnamefont {E.}~\bibnamefont {van Heumen}},
		\bibinfo {author} {\bibfnamefont {J.~G.}\ \bibnamefont {Ram{\'i}rez}},
		\bibinfo {author} {\bibfnamefont {S.}~\bibnamefont {Wang}}, \bibinfo {author}
		{\bibfnamefont {T.}~\bibnamefont {Saerbeck}}, \bibinfo {author}
		{\bibfnamefont {S.}~\bibnamefont {Gu{\'e}non}}, \bibinfo {author}
		{\bibfnamefont {M.}~\bibnamefont {Goldflam}}, \bibinfo {author}
		{\bibfnamefont {L.}~\bibnamefont {Anderegg}}, \bibinfo {author}
		{\bibfnamefont {P.}~\bibnamefont {Kelly}}, \bibinfo {author} {\bibfnamefont
			{A.}~\bibnamefont {Mueller}}, \bibinfo {author} {\bibfnamefont {M.~K.}\
			\bibnamefont {Liu}}, \bibinfo {author} {\bibfnamefont {I.~K.}\ \bibnamefont
			{Schuller}},\ and\ \bibinfo {author} {\bibfnamefont {D.~N.}\ \bibnamefont
			{Basov}},\ }\bibfield  {title} {\bibinfo {title} {Nanotextured {Phase
				Coexistence} in the {Correlated Insulator} {{V$_2$O$_3$}}},\ }\href
	{https://doi.org/10.1038/nphys3882} {\bibfield  {journal} {\bibinfo
			{journal} {Nature Physics}\ }\textbf {\bibinfo {volume} {13}},\ \bibinfo
		{pages} {80} (\bibinfo {year} {2016})}\BibitemShut {NoStop}%
	\bibitem [{\citenamefont {Schindelin}\ \emph {et~al.}(2012)\citenamefont
		{Schindelin}, \citenamefont {Arganda-Carreras}, \citenamefont {Frise},
		\citenamefont {Kaynig}, \citenamefont {Longair}, \citenamefont {Pietzsch},
		\citenamefont {Preibisch}, \citenamefont {Rueden}, \citenamefont {Saalfeld},
		\citenamefont {Schmid}, \citenamefont {Tinevez}, \citenamefont {White},
		\citenamefont {Hartenstein}, \citenamefont {Eliceiri}, \citenamefont
		{Tomancak},\ and\ \citenamefont {Cardona}}]{schindelin2012}%
	\BibitemOpen
	\bibfield  {author} {\bibinfo {author} {\bibfnamefont {J.}~\bibnamefont
			{Schindelin}}, \bibinfo {author} {\bibfnamefont {I.}~\bibnamefont
			{Arganda-Carreras}}, \bibinfo {author} {\bibfnamefont {E.}~\bibnamefont
			{Frise}}, \bibinfo {author} {\bibfnamefont {V.}~\bibnamefont {Kaynig}},
		\bibinfo {author} {\bibfnamefont {M.}~\bibnamefont {Longair}}, \bibinfo
		{author} {\bibfnamefont {T.}~\bibnamefont {Pietzsch}}, \bibinfo {author}
		{\bibfnamefont {S.}~\bibnamefont {Preibisch}}, \bibinfo {author}
		{\bibfnamefont {C.}~\bibnamefont {Rueden}}, \bibinfo {author} {\bibfnamefont
			{S.}~\bibnamefont {Saalfeld}}, \bibinfo {author} {\bibfnamefont
			{B.}~\bibnamefont {Schmid}}, \bibinfo {author} {\bibfnamefont {J.-Y.}\
			\bibnamefont {Tinevez}}, \bibinfo {author} {\bibfnamefont {D.~J.}\
			\bibnamefont {White}}, \bibinfo {author} {\bibfnamefont {V.}~\bibnamefont
			{Hartenstein}}, \bibinfo {author} {\bibfnamefont {K.}~\bibnamefont
			{Eliceiri}}, \bibinfo {author} {\bibfnamefont {P.}~\bibnamefont {Tomancak}},\
		and\ \bibinfo {author} {\bibfnamefont {A.}~\bibnamefont {Cardona}},\
	}\bibfield  {title} {\bibinfo {title} {Fiji: an open-source platform for
			biological-image analysis},\ }\href {https://doi.org/10.1038/nmeth.2019}
	{\bibfield  {journal} {\bibinfo  {journal} {Nature Methods}\ }\textbf
		{\bibinfo {volume} {9}},\ \bibinfo {pages} {676} (\bibinfo {year}
		{2012})}\BibitemShut {NoStop}%
	\bibitem [{\citenamefont {Castellani}\ \emph {et~al.}(1979)\citenamefont
		{Castellani}, \citenamefont {Castro}, \citenamefont {Feinberg},\ and\
		\citenamefont {Ranninger}}]{castellani1979}%
	\BibitemOpen
	\bibfield  {author} {\bibinfo {author} {\bibfnamefont {C.}~\bibnamefont
			{Castellani}}, \bibinfo {author} {\bibfnamefont {C.~D.}\ \bibnamefont
			{Castro}}, \bibinfo {author} {\bibfnamefont {D.}~\bibnamefont {Feinberg}},\
		and\ \bibinfo {author} {\bibfnamefont {J.}~\bibnamefont {Ranninger}},\
	}\bibfield  {title} {\bibinfo {title} {New {{Model Hamiltonian}} for the
			{{Metal-Insulator Transition}}},\ }\href
	{https://doi.org/10.1103/PhysRevLett.43.1957} {\bibfield  {journal} {\bibinfo
			{journal} {Physical Review Letters}\ }\textbf {\bibinfo {volume} {43}},\
		\bibinfo {pages} {1957} (\bibinfo {year} {1979})}\BibitemShut {NoStop}%
	\bibitem [{\citenamefont {Kotliar}\ \emph {et~al.}(2000)\citenamefont
		{Kotliar}, \citenamefont {Lange},\ and\ \citenamefont
		{Rozenberg}}]{kotliar2000}%
	\BibitemOpen
	\bibfield  {author} {\bibinfo {author} {\bibfnamefont {G.}~\bibnamefont
			{Kotliar}}, \bibinfo {author} {\bibfnamefont {E.}~\bibnamefont {Lange}},\
		and\ \bibinfo {author} {\bibfnamefont {M.~J.}\ \bibnamefont {Rozenberg}},\
	}\bibfield  {title} {\bibinfo {title} {Landau {{Theory}} of the {{Finite
					Temperature Mott Transition}}},\ }\href
	{https://doi.org/10.1103/PhysRevLett.84.5180} {\bibfield  {journal} {\bibinfo
			{journal} {Physical Review Letters}\ }\textbf {\bibinfo {volume} {84}},\
		\bibinfo {pages} {5180} (\bibinfo {year} {2000})}\BibitemShut {NoStop}%
	\bibitem [{\citenamefont {D{\'i}az}\ \emph {et~al.}(2023)\citenamefont
		{D{\'i}az}, \citenamefont {Han},\ and\ \citenamefont {Aron}}]{diaz2023}%
	\BibitemOpen
	\bibfield  {author} {\bibinfo {author} {\bibfnamefont {M.~I.}\ \bibnamefont
			{D{\'i}az}}, \bibinfo {author} {\bibfnamefont {J.~E.}\ \bibnamefont {Han}},\
		and\ \bibinfo {author} {\bibfnamefont {C.}~\bibnamefont {Aron}},\ }\bibfield
	{title} {\bibinfo {title} {Electrically {Driven} {Insulator}-to-{Metal}
			{Transition} in a {Correlated} {Insulator}: {{Electronic}} {Mechanism} and
			{Thermal} {Description}},\ }\href
	{https://doi.org/10.1103/PhysRevB.107.195148} {\bibfield  {journal} {\bibinfo
			{journal} {Physical Review B}\ }\textbf {\bibinfo {volume} {107}},\ \bibinfo
		{pages} {195148} (\bibinfo {year} {2023})}\BibitemShut {NoStop}%
	\bibitem [{\citenamefont {Garcke}\ and\ \citenamefont
		{Weikard}(2005)}]{garcke2005a}%
	\BibitemOpen
	\bibfield  {author} {\bibinfo {author} {\bibfnamefont {H.}~\bibnamefont
			{Garcke}}\ and\ \bibinfo {author} {\bibfnamefont {U.}~\bibnamefont
			{Weikard}},\ }\bibfield  {title} {\bibinfo {title} {Numerical {Approximation}
			of the {{Cahn-Larch\'e}} {Equation}},\ }\href
	{https://doi.org/10.1007/s00211-004-0578-x} {\bibfield  {journal} {\bibinfo
			{journal} {Numerische Mathematik}\ }\textbf {\bibinfo {volume} {100}},\
		\bibinfo {pages} {639} (\bibinfo {year} {2005})}\BibitemShut {NoStop}%
	\bibitem [{\citenamefont {Bar}\ \emph {et~al.}(2018)\citenamefont {Bar},
		\citenamefont {Choudhary}, \citenamefont {Ashraf}, \citenamefont {Sujith},
		\citenamefont {Puri}, \citenamefont {Raj},\ and\ \citenamefont
		{Bansal}}]{bar2018}%
	\BibitemOpen
	\bibfield  {author} {\bibinfo {author} {\bibfnamefont {T.}~\bibnamefont
			{Bar}}, \bibinfo {author} {\bibfnamefont {S.~K.}\ \bibnamefont {Choudhary}},
		\bibinfo {author} {\bibfnamefont {M.~A.}\ \bibnamefont {Ashraf}}, \bibinfo
		{author} {\bibfnamefont {K.}~\bibnamefont {Sujith}}, \bibinfo {author}
		{\bibfnamefont {S.}~\bibnamefont {Puri}}, \bibinfo {author} {\bibfnamefont
			{S.}~\bibnamefont {Raj}},\ and\ \bibinfo {author} {\bibfnamefont
			{B.}~\bibnamefont {Bansal}},\ }\bibfield  {title} {\bibinfo {title} {Kinetic
			{Spinodal} {Instabilities} in the {Mott} {Transition} in {V$_2$O$_3$}:
			{Evidence} from {Hysteresis} {Scaling} and {Dissipative} {Phase}
			{Ordering}},\ }\href {https://doi.org/10.1103/physrevlett.121.045701}
	{\bibfield  {journal} {\bibinfo  {journal} {Physical Review Letters}\
		}\textbf {\bibinfo {volume} {121}},\ \bibinfo {pages} {045701} (\bibinfo
		{year} {2018})}\BibitemShut {NoStop}%
	\bibitem [{\citenamefont {Kundu}\ \emph {et~al.}(2020)\citenamefont {Kundu},
		\citenamefont {Bar}, \citenamefont {Nayak},\ and\ \citenamefont
		{Bansal}}]{kundu2020}%
	\BibitemOpen
	\bibfield  {author} {\bibinfo {author} {\bibfnamefont {S.}~\bibnamefont
			{Kundu}}, \bibinfo {author} {\bibfnamefont {T.}~\bibnamefont {Bar}}, \bibinfo
		{author} {\bibfnamefont {R.~K.}\ \bibnamefont {Nayak}},\ and\ \bibinfo
		{author} {\bibfnamefont {B.}~\bibnamefont {Bansal}},\ }\bibfield  {title}
	{\bibinfo {title} {Critical {Slowing} {Down} at the {Abrupt} {Mott}
			{Transition}: {When} the {First}-{Order} {Phase} {Transition} {Becomes}
			{Zeroth} {Order} and {Looks} {Like} {Second} {Order}},\ }\href
	{https://doi.org/10.1103/physrevlett.124.095703} {\bibfield  {journal}
		{\bibinfo  {journal} {Physical Review Letters}\ }\textbf {\bibinfo {volume}
			{124}},\ \bibinfo {pages} {095703} (\bibinfo {year} {2020})}\BibitemShut
	{NoStop}%
	\bibitem [{\citenamefont {Ronchi}\ \emph {et~al.}(2019)\citenamefont {Ronchi},
		\citenamefont {Homm}, \citenamefont {Menghini}, \citenamefont {Franceschini},
		\citenamefont {Maccherozzi}, \citenamefont {Banfi}, \citenamefont {Ferrini},
		\citenamefont {Cilento}, \citenamefont {Parmigiani}, \citenamefont {Dhesi},
		\citenamefont {Fabrizio}, \citenamefont {Locquet},\ and\ \citenamefont
		{Giannetti}}]{ronchi2019}%
	\BibitemOpen
	\bibfield  {author} {\bibinfo {author} {\bibfnamefont {A.}~\bibnamefont
			{Ronchi}}, \bibinfo {author} {\bibfnamefont {P.}~\bibnamefont {Homm}},
		\bibinfo {author} {\bibfnamefont {M.}~\bibnamefont {Menghini}}, \bibinfo
		{author} {\bibfnamefont {P.}~\bibnamefont {Franceschini}}, \bibinfo {author}
		{\bibfnamefont {F.}~\bibnamefont {Maccherozzi}}, \bibinfo {author}
		{\bibfnamefont {F.}~\bibnamefont {Banfi}}, \bibinfo {author} {\bibfnamefont
			{G.}~\bibnamefont {Ferrini}}, \bibinfo {author} {\bibfnamefont
			{F.}~\bibnamefont {Cilento}}, \bibinfo {author} {\bibfnamefont
			{F.}~\bibnamefont {Parmigiani}}, \bibinfo {author} {\bibfnamefont {S.~S.}\
			\bibnamefont {Dhesi}}, \bibinfo {author} {\bibfnamefont {M.}~\bibnamefont
			{Fabrizio}}, \bibinfo {author} {\bibfnamefont {J.-P.}\ \bibnamefont
			{Locquet}},\ and\ \bibinfo {author} {\bibfnamefont {C.}~\bibnamefont
			{Giannetti}},\ }\bibfield  {title} {\bibinfo {title} {Early-{Stage}
			{Dynamics} of {Metallic} {Droplets} {Embedded} in the {Nanotextured} {{Mott}}
			{Insulating} {Phase} of {V$_2$O$_3$}},\ }\href
	{https://doi.org/10.1103/PhysRevB.100.075111} {\bibfield  {journal} {\bibinfo
			{journal} {Physical Review B}\ }\textbf {\bibinfo {volume} {100}},\ \bibinfo
		{pages} {075111} (\bibinfo {year} {2019})}\BibitemShut {NoStop}%
	\bibitem [{\citenamefont {Giorgianni}\ \emph {et~al.}(2019)\citenamefont
		{Giorgianni}, \citenamefont {Sakai},\ and\ \citenamefont
		{Lupi}}]{giorgianni2019}%
	\BibitemOpen
	\bibfield  {author} {\bibinfo {author} {\bibfnamefont {F.}~\bibnamefont
			{Giorgianni}}, \bibinfo {author} {\bibfnamefont {J.}~\bibnamefont {Sakai}},\
		and\ \bibinfo {author} {\bibfnamefont {S.}~\bibnamefont {Lupi}},\ }\bibfield
	{title} {\bibinfo {title} {Overcoming the thermal regime for the
			electric-field driven {Mott} transition in vanadium sesquioxide},\ }\href
	{https://doi.org/10.1038/s41467-019-09137-6} {\bibfield  {journal} {\bibinfo
			{journal} {Nature Communications}\ }\textbf {\bibinfo {volume} {10}},\
		\bibinfo {pages} {1159} (\bibinfo {year} {2019})}\BibitemShut {NoStop}%
	\bibitem [{\citenamefont {Cavalleri}\ \emph {et~al.}(2001)\citenamefont
		{Cavalleri}, \citenamefont {Tóth}, \citenamefont {Siders}, \citenamefont
		{Squier}, \citenamefont {Ráksi}, \citenamefont {Forget},\ and\ \citenamefont
		{Kieffer}}]{cavalleri2001}%
	\BibitemOpen
	\bibfield  {author} {\bibinfo {author} {\bibfnamefont {A.}~\bibnamefont
			{Cavalleri}}, \bibinfo {author} {\bibfnamefont {C.}~\bibnamefont {Tóth}},
		\bibinfo {author} {\bibfnamefont {C.~W.}\ \bibnamefont {Siders}}, \bibinfo
		{author} {\bibfnamefont {J.~A.}\ \bibnamefont {Squier}}, \bibinfo {author}
		{\bibfnamefont {F.}~\bibnamefont {Ráksi}}, \bibinfo {author} {\bibfnamefont
			{P.}~\bibnamefont {Forget}},\ and\ \bibinfo {author} {\bibfnamefont {J.~C.}\
			\bibnamefont {Kieffer}},\ }\bibfield  {title} {\bibinfo {title} {Femtosecond
			{Structural} {Dynamics} in {VO$_2$} during an {Ultrafast} {Solid}-{Solid}
			{Phase} {Transition}},\ }\href
	{https://doi.org/10.1103/physrevlett.87.237401} {\bibfield  {journal}
		{\bibinfo  {journal} {Physical Review Letters}\ }\textbf {\bibinfo {volume}
			{87}},\ \bibinfo {pages} {237401} (\bibinfo {year} {2001})}\BibitemShut
	{NoStop}%
	\bibitem [{\citenamefont {Wall}\ \emph {et~al.}(2018)\citenamefont {Wall},
		\citenamefont {Yang}, \citenamefont {Vidas}, \citenamefont {Chollet},
		\citenamefont {Glownia}, \citenamefont {Kozina}, \citenamefont {Katayama},
		\citenamefont {Henighan}, \citenamefont {Jiang}, \citenamefont {Miller},
		\citenamefont {Reis}, \citenamefont {Boatner}, \citenamefont {Delaire},\ and\
		\citenamefont {Trigo}}]{wall2018}%
	\BibitemOpen
	\bibfield  {author} {\bibinfo {author} {\bibfnamefont {S.}~\bibnamefont
			{Wall}}, \bibinfo {author} {\bibfnamefont {S.}~\bibnamefont {Yang}}, \bibinfo
		{author} {\bibfnamefont {L.}~\bibnamefont {Vidas}}, \bibinfo {author}
		{\bibfnamefont {M.}~\bibnamefont {Chollet}}, \bibinfo {author} {\bibfnamefont
			{J.~M.}\ \bibnamefont {Glownia}}, \bibinfo {author} {\bibfnamefont
			{M.}~\bibnamefont {Kozina}}, \bibinfo {author} {\bibfnamefont
			{T.}~\bibnamefont {Katayama}}, \bibinfo {author} {\bibfnamefont
			{T.}~\bibnamefont {Henighan}}, \bibinfo {author} {\bibfnamefont
			{M.}~\bibnamefont {Jiang}}, \bibinfo {author} {\bibfnamefont {T.~A.}\
			\bibnamefont {Miller}}, \bibinfo {author} {\bibfnamefont {D.~A.}\
			\bibnamefont {Reis}}, \bibinfo {author} {\bibfnamefont {L.~A.}\ \bibnamefont
			{Boatner}}, \bibinfo {author} {\bibfnamefont {O.}~\bibnamefont {Delaire}},\
		and\ \bibinfo {author} {\bibfnamefont {M.}~\bibnamefont {Trigo}},\ }\bibfield
	{title} {\bibinfo {title} {Ultrafast disordering of vanadium dimers in
			photoexcited {VO$_2$}},\ }\href {https://doi.org/10.1126/science.aau3873}
	{\bibfield  {journal} {\bibinfo  {journal} {Science}\ }\textbf {\bibinfo
			{volume} {362}},\ \bibinfo {pages} {572} (\bibinfo {year}
		{2018})}\BibitemShut {NoStop}%
	\bibitem [{\citenamefont {Otto}\ \emph {et~al.}(2018)\citenamefont {Otto},
		\citenamefont {René~de Cotret}, \citenamefont {Valverde-Chavez},
		\citenamefont {Tiwari}, \citenamefont {Émond}, \citenamefont {Chaker},
		\citenamefont {Cooke},\ and\ \citenamefont {Siwick}}]{otto2018}%
	\BibitemOpen
	\bibfield  {author} {\bibinfo {author} {\bibfnamefont {M.~R.}\ \bibnamefont
			{Otto}}, \bibinfo {author} {\bibfnamefont {L.~P.}\ \bibnamefont {René~de
				Cotret}}, \bibinfo {author} {\bibfnamefont {D.~A.}\ \bibnamefont
			{Valverde-Chavez}}, \bibinfo {author} {\bibfnamefont {K.~L.}\ \bibnamefont
			{Tiwari}}, \bibinfo {author} {\bibfnamefont {N.}~\bibnamefont {Émond}},
		\bibinfo {author} {\bibfnamefont {M.}~\bibnamefont {Chaker}}, \bibinfo
		{author} {\bibfnamefont {D.~G.}\ \bibnamefont {Cooke}},\ and\ \bibinfo
		{author} {\bibfnamefont {B.~J.}\ \bibnamefont {Siwick}},\ }\bibfield  {title}
	{\bibinfo {title} {How optical excitation controls the structure and
			properties of vanadium dioxide},\ }\href
	{https://doi.org/10.1073/pnas.1808414115} {\bibfield  {journal} {\bibinfo
			{journal} {Proceedings of the National Academy of Sciences}\ }\textbf
		{\bibinfo {volume} {116}},\ \bibinfo {pages} {450} (\bibinfo {year}
		{2018})}\BibitemShut {NoStop}%
\end{thebibliography}

\begin{thebibliography}{5}%
	\makeatletter
	\providecommand \@ifxundefined [1]{%
		\@ifx{#1\undefined}
	}%
	\providecommand \@ifnum [1]{%
		\ifnum #1\expandafter \@firstoftwo
		\else \expandafter \@secondoftwo
		\fi
	}%
	\providecommand \@ifx [1]{%
		\ifx #1\expandafter \@firstoftwo
		\else \expandafter \@secondoftwo
		\fi
	}%
	\providecommand \natexlab [1]{#1}%
	\providecommand \enquote  [1]{``#1''}%
	\providecommand \bibnamefont  [1]{#1}%
	\providecommand \bibfnamefont [1]{#1}%
	\providecommand \citenamefont [1]{#1}%
	\providecommand \href@noop [0]{\@secondoftwo}%
	\providecommand \href [0]{\begingroup \@sanitize@url \@href}%
	\providecommand \@href[1]{\@@startlink{#1}\@@href}%
	\providecommand \@@href[1]{\endgroup#1\@@endlink}%
	\providecommand \@sanitize@url [0]{\catcode `\\12\catcode `\$12\catcode
		`\&12\catcode `\#12\catcode `\^12\catcode `\_12\catcode `\%12\relax}%
	\providecommand \@@startlink[1]{}%
	\providecommand \@@endlink[0]{}%
	\providecommand \url  [0]{\begingroup\@sanitize@url \@url }%
	\providecommand \@url [1]{\endgroup\@href {#1}{\urlprefix }}%
	\providecommand \urlprefix  [0]{URL }%
	\providecommand \Eprint [0]{\href }%
	\providecommand \doibase [0]{https://doi.org/}%
	\providecommand \selectlanguage [0]{\@gobble}%
	\providecommand \bibinfo  [0]{\@secondoftwo}%
	\providecommand \bibfield  [0]{\@secondoftwo}%
	\providecommand \translation [1]{[#1]}%
	\providecommand \BibitemOpen [0]{}%
	\providecommand \bibitemStop [0]{}%
	\providecommand \bibitemNoStop [0]{.\EOS\space}%
	\providecommand \EOS [0]{\spacefactor3000\relax}%
	\providecommand \BibitemShut  [1]{\csname bibitem#1\endcsname}%
	\let\auto@bib@innerbib\@empty
	%</preamble>
	
	
\let\addcontentsline\oldaddcontentsline
	
	\section{References}	
	
	
	\bibitem [{\citenamefont {Lange}\ \emph {et~al.}(2017)\citenamefont {Lange},
		\citenamefont {Gu{\'e}non}, \citenamefont {Lever}, \citenamefont {Kleiner},\
		and\ \citenamefont {Koelle}}]{lange2017s}%
	\BibitemOpen
	\bibfield  {author} {\bibinfo {author} {\bibfnamefont {M.~M.}\ \bibnamefont
			{Lange}}, \bibinfo {author} {\bibfnamefont {S.}~\bibnamefont {Gu{\'e}non}},
		\bibinfo {author} {\bibfnamefont {F.}~\bibnamefont {Lever}}, \bibinfo
		{author} {\bibfnamefont {R.}~\bibnamefont {Kleiner}},\ and\ \bibinfo {author}
		{\bibfnamefont {D.}~\bibnamefont {Koelle}},\ }\bibfield  {title} {\bibinfo
		{title} {A {High}-{Resolution} {Combined} {Scanning} {Laser} and {Widefield}
			{Polarizing} {Microscope} for {Imaging} at {Temperatures} from 4 {{K}} to 300
			{{K}}},\ }\href {https://doi.org/10.1063/1.5009529} {\bibfield  {journal}
		{\bibinfo  {journal} {Review of Scientific Instruments}\ }\textbf {\bibinfo
			{volume} {88}},\ \bibinfo {pages} {123705} (\bibinfo {year}
		{2017})}\BibitemShut {NoStop}%
	\bibitem [{\citenamefont {Lange}(2018)}]{lange2018s}%
	\BibitemOpen
	\bibfield  {author} {\bibinfo {author} {\bibfnamefont {M.~M.}\ \bibnamefont
			{Lange}},\ }\href@noop {} {\bibinfo {title} {A {High-Resolution Polarizing
				Microscope} for {Cryogenic} {Imaging}: {{Development}} and {{Application}} to
			{{Investigations}} on {{Twin Walls}} in {{SrTiO$_3$}} and the
			{{Metal-Insulator Transition}} in {{V$_2$O$_3$}}}},\ \bibinfo {howpublished}
	{Ph.D. thesis, Universität Tübingen} (\bibinfo {year} {2018})\BibitemShut
	{NoStop}%
	\bibitem [{\citenamefont {Trastoy}\ \emph {et~al.}(2018)\citenamefont
		{Trastoy}, \citenamefont {Kalcheim}, \citenamefont {{del Valle}},
		\citenamefont {Valmianski},\ and\ \citenamefont {Schuller}}]{trastoy2018a}%
	\BibitemOpen
	\bibfield  {author} {\bibinfo {author} {\bibfnamefont {J.}~\bibnamefont
			{Trastoy}}, \bibinfo {author} {\bibfnamefont {Y.}~\bibnamefont {Kalcheim}},
		\bibinfo {author} {\bibfnamefont {J.}~\bibnamefont {{del Valle}}}, \bibinfo
		{author} {\bibfnamefont {I.}~\bibnamefont {Valmianski}},\ and\ \bibinfo
		{author} {\bibfnamefont {I.~K.}\ \bibnamefont {Schuller}},\ }\bibfield
	{title} {\bibinfo {title} {Enhanced {Metal}\textendash {Insulator}
			{Transition} in {{V$_2$O$_3$}} by {Thermal} {Quenching} after {Growth}},\
	}\href {https://doi.org/10.1007/s10853-018-2214-7} {\bibfield  {journal}
		{\bibinfo  {journal} {Journal of Materials Science}\ }\textbf {\bibinfo
			{volume} {53}},\ \bibinfo {pages} {9131} (\bibinfo {year}
		{2018})}\BibitemShut {NoStop}%
	\bibitem [{\citenamefont {del Valle}\ \emph {et~al.}(2019)\citenamefont {del
			Valle}, \citenamefont {Salev}, \citenamefont {Tesler}, \citenamefont
		{Vargas}, \citenamefont {Kalcheim}, \citenamefont {Wang}, \citenamefont
		{Trastoy}, \citenamefont {Lee}, \citenamefont {Kassabian}, \citenamefont
		{Ramírez}, \citenamefont {Rozenberg},\ and\ \citenamefont
		{Schuller}}]{valle2019}%
	\BibitemOpen
	\bibfield  {author} {\bibinfo {author} {\bibfnamefont {J.}~\bibnamefont {del
				Valle}}, \bibinfo {author} {\bibfnamefont {P.}~\bibnamefont {Salev}},
		\bibinfo {author} {\bibfnamefont {F.}~\bibnamefont {Tesler}}, \bibinfo
		{author} {\bibfnamefont {N.~M.}\ \bibnamefont {Vargas}}, \bibinfo {author}
		{\bibfnamefont {Y.}~\bibnamefont {Kalcheim}}, \bibinfo {author}
		{\bibfnamefont {P.}~\bibnamefont {Wang}}, \bibinfo {author} {\bibfnamefont
			{J.}~\bibnamefont {Trastoy}}, \bibinfo {author} {\bibfnamefont {M.-H.}\
			\bibnamefont {Lee}}, \bibinfo {author} {\bibfnamefont {G.}~\bibnamefont
			{Kassabian}}, \bibinfo {author} {\bibfnamefont {J.~G.}\ \bibnamefont
			{Ramírez}}, \bibinfo {author} {\bibfnamefont {M.~J.}\ \bibnamefont
			{Rozenberg}},\ and\ \bibinfo {author} {\bibfnamefont {I.~K.}\ \bibnamefont
			{Schuller}},\ }\bibfield  {title} {\bibinfo {title} {Subthreshold firing in
			{M}ott nanodevices},\ }\href {https://doi.org/10.1038/s41586-019-1159-6}
	{\bibfield  {journal} {\bibinfo  {journal} {Nature}\ }\textbf {\bibinfo
			{volume} {569}},\ \bibinfo {pages} {388} (\bibinfo {year}
		{2019})}\BibitemShut {NoStop}%
	\bibitem [{\citenamefont {Stoliar}\ \emph {et~al.}(2017)\citenamefont
		{Stoliar}, \citenamefont {Tranchant}, \citenamefont {Corraze}, \citenamefont
		{Janod}, \citenamefont {Besland}, \citenamefont {Tesler}, \citenamefont
		{Rozenberg},\ and\ \citenamefont {Cario}}]{stoliar2017}%
	\BibitemOpen
	\bibfield  {author} {\bibinfo {author} {\bibfnamefont {P.}~\bibnamefont
			{Stoliar}}, \bibinfo {author} {\bibfnamefont {J.}~\bibnamefont {Tranchant}},
		\bibinfo {author} {\bibfnamefont {B.}~\bibnamefont {Corraze}}, \bibinfo
		{author} {\bibfnamefont {E.}~\bibnamefont {Janod}}, \bibinfo {author}
		{\bibfnamefont {M.}~\bibnamefont {Besland}}, \bibinfo {author} {\bibfnamefont
			{F.}~\bibnamefont {Tesler}}, \bibinfo {author} {\bibfnamefont
			{M.}~\bibnamefont {Rozenberg}},\ and\ \bibinfo {author} {\bibfnamefont
			{L.}~\bibnamefont {Cario}},\ }\bibfield  {title} {\bibinfo {title} {A
			leaky‐integrate‐and‐fire neuron analog realized with a {M}ott
			insulator},\ }\href {https://doi.org/10.1002/adfm.201604740} {\bibfield
		{journal} {\bibinfo  {journal} {Advanced Functional Materials}\ }\textbf
		{\bibinfo {volume} {27}},\ \bibinfo {pages} {1604740} (\bibinfo {year}
		{2017})}\BibitemShut {NoStop}%
\end{thebibliography}
\end{document}